\documentclass[twocolumn]{aastex631}
\usepackage{graphicx}
\usepackage{xcolor}
\usepackage{listings}
\usepackage{float}
\usepackage[caption=false]{subfig}
\newcommand{\nprofit}{{\sc nProFit}}
\shorttitle{{\sc nProFit}: a tool for dynamical models fitting}
\shortauthors{Cuevas-Otahola et al.}
\graphicspath{{./}{figures/}}

\usepackage{float}
\usepackage{mathtools}
\usepackage{subfig}

\begin{document}

\title{{\sc nProFit}: a tool for fitting the surface brightness profiles of star clusters with dynamical models}

\author[0000-0002-1046-1500]{B. Cuevas-Otahola}
\author{Y. D. Mayya}
\author{I. Puerari}
\author{D. Rosa-Gonz\'alez}
\affiliation{Instituto Nacional de Astrof\'isica, \'Optica y Electr\'onica, 72840 Puebla, Mexico \\}




\begin{abstract}
The surface brightness profiles (SBPs) of star clusters hold invaluable information on the dynamical state of clusters. The observed SBPs of star clusters, especially that of globular clusters, are in good agreement with the SBPs expected for isothermal spheres containing stars of reduced kinetic energies. However, the SBPs of configurations that satisfy these theoretical criteria cannot be uniquely  expressed by analytical formulae, which had hindered the analysis of dynamical state of observed clusters in external galaxies. To counter this shortcoming, it has become a practice to use empirical fitting formulae that best represent the core and halo characteristics of theoretical models. We here present a general purpose code, named {\sc nProFit}, that allows fitting of the surface brightness profiles of extragalactic star clusters to theoretical star clusters, defined by dynamical models of \citet{King_dyn} and \citet{Wilson_dyn}. In addition, we also incorporated theoretical models that result in power-law surface brightness profiles represented by \citet{Elson}. The code returns the basic size parameters such as core radius, half-light radius and tidal radius, as well as dynamically relevant parameters, such as the volume and surface density profiles, velocity dispersion profile, total mass and the binding energy for a user-fixed  mass-to-light ratio. 
The usefulness of the code in the dynamical study of extragalactic clusters has been already illustrated in \citet{Cuevas2020}. The code, which is python-based at the user end, but makes calls to advanced routines in Pyraf and Fortran, is now available for public use. We provide example scripts and mock clusters in the installation package as guide to users.
\end{abstract}



\section{Introduction}
Structural parameters of star clusters serve as proxies to give insights on their dynamical evolutionary state.  These parameters, namely, core radius, half-mass and half-light radius, tidal radius and concentration index, can be obtained by fitting theoretical intensity profiles to the surface brightness profiles (SBP) of clusters.  For instance, the \citet{King_emp} is one of the most widely-used profiles to obtain the structural parameters of old clusters, such as globular clusters (GCs). In the pioneering work, \citet{King_dyn} demonstrated that the observed form of the profiles of GCs belong to the family of the surface mass density profiles corresponding to self-gravitating isothermal spheres of lower kinetic energies. \citet{Wilson_dyn}  proposed a dynamical profile resembling the structure of a \citet{King_dyn} profile, with a larger halo, in order to fit the  SBP of elliptical galaxies.   Years later,  \citet{Elson} found that the SBPs of clusters in the Large Magellanic Cloud (LMC) do not have a noticeable break corresponding to tidal radius of King profiles. They found their profiles are better fit by power-law functions  rather than King profiles. In these power-law profiles, the shapes of the extended haloes are characterised by $\gamma$, with $\gamma=$2 corresponding to the profiles of infinite mass
isothermal models. Profiles with $\gamma>$2 are steeper and have finite masses.

Over the last two decades, the observed SBPs of extragalactic star clusters have been analysed in several studies to obtain structural parameters using theoretical profiles. The most frequently used tools in these studies are {\sc ishape}  \citep{Larsenishape} and {\sc galfit} \citep{Penggalfit}, both of which are available for public use. Both
these tools fit two-dimensional empirical profiles to the sky-subtracted observed 2D images of the objects under study.  Geometrical parameters such as position angle and ellipticities are fitted as well. These codes are mainly used to obtain the size parameters such as core radius, half-light radius ($R_{\rm h}$) and tidal radius for assumed shape of the profile. This is the best one can hope to do in images where star clusters are only marginally resolved and the background subtraction errors do not permit an analysis of the shapes of profiles in their external parts. 

One of the defining parameters of star clusters is their mass, which is determined either using photometric techniques, or using dynamical models. The photometric mass is routinely determined using the observed luminosity along with a value for the mass-to-light ratio appropriate to the population of stars in the cluster. The mass-to-light ratio is calculated in population synthesis models, and is a function of age and metallicity \citep{BruzualCharlot2003}. On the other hand, the dynamical mass is based on the determination of motions of stars under the influence of the collective force of all its stars, and is defined as the Virial mass,  $M_{\rm vir} = \eta \sigma_{\rm p}^2 R_{\rm h}/G$, where  $\sigma_{\rm p}$ is the velocity dispersion projected along the line of sight, and $\eta$ is the Virial factor which depends on the shape of the profile \citep{GielesSana2010}. The photometric and dynamical masses do not always agree for clusters for which both the measurements are available \citep[see e.g.][]{McLaughlin2008, GielesSana2010}. Almost a factor of 10 uncertainty in  $\eta$ is one of the sources of the disagreements between the photometric and dynamical masses. This uncertainty can be avoided by theoretically calculating the radial profiles of dispersion velocities for the model that fits the observed SBP. Such calculations can be carried out for profiles that have an underlying physical model as illustrated by \citet{Barmby2007} and \citet{McLaughlin2008}, who characterized the SBPs of globular clusters in M31 and NGC5128, respectively. The recent availability of high resolution multi-object or integral field unit-fed spectrographs on large telescopes \citep[e.g.][]{GildePaz2018}
make it possible the determination of $\sigma_{\rm p}$ of large samples of star clusters in nearby galaxies, which calls for the recovery of the $\sigma_{\rm p}$ corresponding to the best-fit models.

The Hubble Space Telescope (HST) images of nearby galaxies (distance $\lesssim$5~Mpc) contain star clusters whose profiles are good enough for a characterisation of the shape of the outer halo, using physical models. However, the absence of a publicly available code is a handicap to analyse the profiles of star clusters in these images. The purpose of the present work is to develop a user-friendly code that can analyse the SBPs of star clusters on the HST images to obtain simultaneously the core and halo parameters, in addition to $\sigma_{\rm p}$. In order to achieve this, we follow the procedure outlined by \citet{Elson}, as well as the prescription to fit dynamical models (King and Wilson) adopted by \citet{McLaughlin2000}, \citet{McLaughlin2005} and \citet{Sollima2015}.

We here introduce {\sc nProFit} (Profile Fitting tool of n-objects)\footnote{{\sc nProFit} is publicly available in the GitHub repository\\ \url{https://github.com/umbramortem/nProFit}} for fitting dynamical models to 1D SBPs for a user-given list of star clusters in a single image. The dynamical models considered are King \citep{King_dyn}, Wilson \citep{Wilson_dyn}, and EFF \citep{Elson}. For the former two models, SBPs are generated for dynamically stable isothermal clusters of reduced kinetic energies, whereas for the latter profile Jeans' equation and the subsequent Poisson's equations are solved to obtain dynamical parameters that are consistent with the observed SBPs. From the structural parameters obtained by {\sc nProFit} (scale radius $r_d$ for EFF or $r_0$ for King and Wilson, and shape parameters $\gamma$ for EFF and $W_0$ for King and Wilson), our proposed tool computes dynamically relevant parameters such as mass, surface and volume mass densities, velocity dispersion and binding energy, as well as tidal radius and core radius.

In \S\ref{Sec:algo}, we describe the {\sc nProFit} underlying algorithm, starting with the program initialization, moving subsequently to the background subtraction, SBP extraction, and subsequently describing the fitted models and the corresponding parameters derivation, following the $\chi^2$ minimization technique. In \S\ref{Sec:nprofit_apps} we introduce the simulation tool {\sc mksample}, and use it to create a mock clusters sample to illustrate the operation of {\sc nProFit}. Finally, in \S\ref{Sec:conclu} we show our conclusions and the future directions for {\sc nProFit}.

\section{The algorithm}{\label{Sec:algo}}

{\sc nProFit} stands for n-Profile Fitting Tool.  This tool was developed to extract and fit the observed SBPs of $n$ objects in an image.  This task is carried out by following a series of steps implemented in {\sc Fortran}, {\sc Python} and {\sc PyRAF} routines. In this section, we describe the structure of our code.

\subsection{The scope}

\nprofit\ is developed to obtain the structural parameters of star clusters on the fits format science images of nearby galaxies taken with the HST, such as those in the ACS Nearby Galaxy Survey Treasury \citep[ANGST,][]{AngstSurvey}, which has more than 60 galaxies at distances $\lesssim$4~Mpc. There are two kinds of star clusters that can be easily detected on the HST images of nearby galaxies --- GCs and Super Star Clusters (SSCs). These clusters typically have a half-light radius less than 10~pc, which allows them to be distinguished from stars on the HST images up to distances $\sim$5~Mpc. The cores of star clusters are supported against the gravity by the pressure exerted by the random motions of stars and hence are well modelled as isothermal spheres of finite kinetic energies. We hence structured our code to fit the observed SBPs with theoretical profiles for families of stable clusters. Theoretical profiles are defined in mass surface density, which is related to the observed SBPs through the mass-to-light ratio, which is assumed to be be 1 $\rm M_\odot/L_\odot$ and independent of radius in this work. For theoretical clusters containing stars of equal mass and without internal dust such as those defined by \citet{King_dyn} models, this is a good assumption. However, real clusters have stars of a range of masses, with a tendency for the most massive stars to sink to the center as the cluster dynamically evolves, which produces a color gradient that is bluer towards the center \citep[see e.g.][]{Djorgovski1993}. Given that the bulk of the mass of a cluster is in low-mass red stars for \citet{KroupaIMF} and other initial mass functions, SBP in a red filter is expected to trace the mass-density profile better than that in a blue filter. Filters at longer wavelengths also are less affected by possible presence of differential reddening due to patchy internal dust \citep{Cardelli1989}. Among the commonly used HST/ACS filters, F814W is the reddest filter, and hence we recommend the use of F814W images for obtaining SBPs.

Morphological analysis of extragalactic star clusters suffers from two problems:  (1) background variation --- star clusters are often encountered in zones with varying background values such as in the spiral arms of galaxies, which requires a measurement of a local background value for each object, and (2) object crowding --- star clusters are hardly isolated. Both these facts affect the surface brightness of the profile in its external parts. We have built-in algorithms in the code to address both these issues.

\subsection{Initialization and data preparation}

The analysis of SBPs of objects depends on a number of data-dependent parameters.
As a first step, the code reads an input ASCII file containing specific analysis parameters. An example of program input can be found in Appendix \ref{Ap:nprofit}. One of the parameters in the input file is the name of a list containing coordinates of $n$ objects. The code obtains 2D sub-images centered on each object in the list. Each sub-image is obtained by trimming the original image of user-defined sizes and centred on the coordinates in the object list, either world coordinates (WCS) or in pixels.  These individual images centred at each object coordinates allow us to study the objects separately, since datasets are typically constituted by several objects. 

\subsection{Background estimation and subtraction}\label{sec:bg_sub}

Estimating and subtracting the background value accurately is crucial in the determination of structural parameters.  An erroneous estimation of the background value may result in an overestimation or underestimation of the derived structural parameters.
We implemented two techniques to estimate the background values locally in each sub-image image, which are explained below.

{\it Statistical determination in the corners:} This strategy is based on the computation of median values in four corners of each sub-image over box sizes of 10\% the size of each sub-image. The use of median, rather than the mean, ensures that each estimated background value is not much affected by any contaminating object in the corners. The availability of four measurements allows us to check the uniformity of the background. The minimum of the four median and root mean square (rms) values are taken as the optimum background and rms values for the object under analysis.

{\it k$\times\sigma$ clipped images:} This method uses the whole sub-image to obtain an optimum background value for the object. A background image is obtained by iteratively rejecting pixels with values above and below $k\times\sigma$ around the median value in user-defined small boxes. The median and rms values of this background image are taken as the optimum background and rms values for the object under analysis. The obtained values are found to stabilise after $\sim$ten clipping iterations with $k=3$ (i.e. the procedure is performed rejecting values above $3\sigma$ to estimate the background value). For this method to obtain reliable background value, the box size should be at least twice the size of the analyzed object.

Both these methods ignore variations in the background value over scales of the sub-image. In principle, gradient in the background values over scales of the cluster size can be obtained by including the background function while fitting the 1D SBPs \citep{MackeyGilmore2003a}.  However, we found that the background value determined by this method does not easily converge to stable values, like the two methods described above. We have used these two techniques to analyse SSCs in M82, and obtained reliable SBPs and therefore accurate structural parameters \citep{Cuevas2020,Cuevas2020b}. The obtained background level values are subtracted to their corresponding 2-D individual images, 
resulting in background-subtracted sub-images. 

\subsection{Masking contaminants}

In some cases, the distance between objects are too small, hindering an accurate profile extraction. Analysis of such clusters requires setting specific constraints on the fitting procedure by either masking the contaminants or setting smaller fitting radii.  

{\sc nProFit} uses the masking capability that the ellipse task provides to obtain 1D SBPs of sources without contribution from close contaminant sources. The mask input file is an ASCII file containing the number of circular masks, along with their spatial coordinates and radii in pixels.

We draw particular attention to cases where a contaminant source is located in the corners of the figure, causing noticeable variations in the background level.  If a contaminant source fits the latter scenario, it will be masked by {\sc nProFit} prior the background level estimation.  

\subsection{Surface brightness profile extraction}

Star clusters on the HST images are not resolved enough to obtain SBPs using the star count method that is normally used to analyse star clusters in the Milky Way and the LMC 
\citep{Brandl2006, MackeyGilmore2003a}. We used the {\sc ellipse} task in the {\sc IRAF/STSDAS} package \citep{Ellipse_iraf1987}, one of the most widely used tools to extract the SBPs from 2D images, for analysing extragalactic star clusters. The {\sc ellipse} task obtains SBPs by azimuthally averaging intensities in elliptical annular zones. The task allows for variation of center, ellipticity ($\epsilon=1-b/a$, where $a$ and $b$ are major and minor axes of the ellipse) and position angle (PA) of the major axis of the successive ellipses. Linear increments in the semi-major axis of the successive ellipses were used to calculate the intensity profiles. We defined concentric ellipses with their centers fixed at the user-supplied object coordinates. The $\epsilon$ and PA were also fixed either to the user-provided values or to their asymptotic values. In the latter case, {\sc nProFit} computes the ellipticity value by running the {\sc ellipse} task without fixing the ellipticity values in a first iteration, and obtains the SBPs in a second run of {\sc ellipse} after fixing the values of $\epsilon$ and PA. In this scenario, {\sc nProFit} analyzes the radial profiles of ellipticities from the first run, setting the ellipticity and PA values as the ones corresponding to the radius at which the ellipticity values are nearly constant.  In the majority of the cases, the value where the ellipticity stabilizes  matches the radius containing half the total cumulative flux, i.e. the effective radius.  During the ellipticity estimation, {\sc nProFit} avoids the most inner radii that does not provide reliable measurements. The {\sc ellipse} task obtains cumulative intensities in addition to the SBPs for every fitted ellipse. The $k-sigma$ clipped rms fluctuations in the azimuthal intensities for each fitted ellipse are taken as the error on the measured intensities. All masked pixels are excluded from the analysis during the ellipse fitting. 

\begin{figure}
\includegraphics[width=\columnwidth]{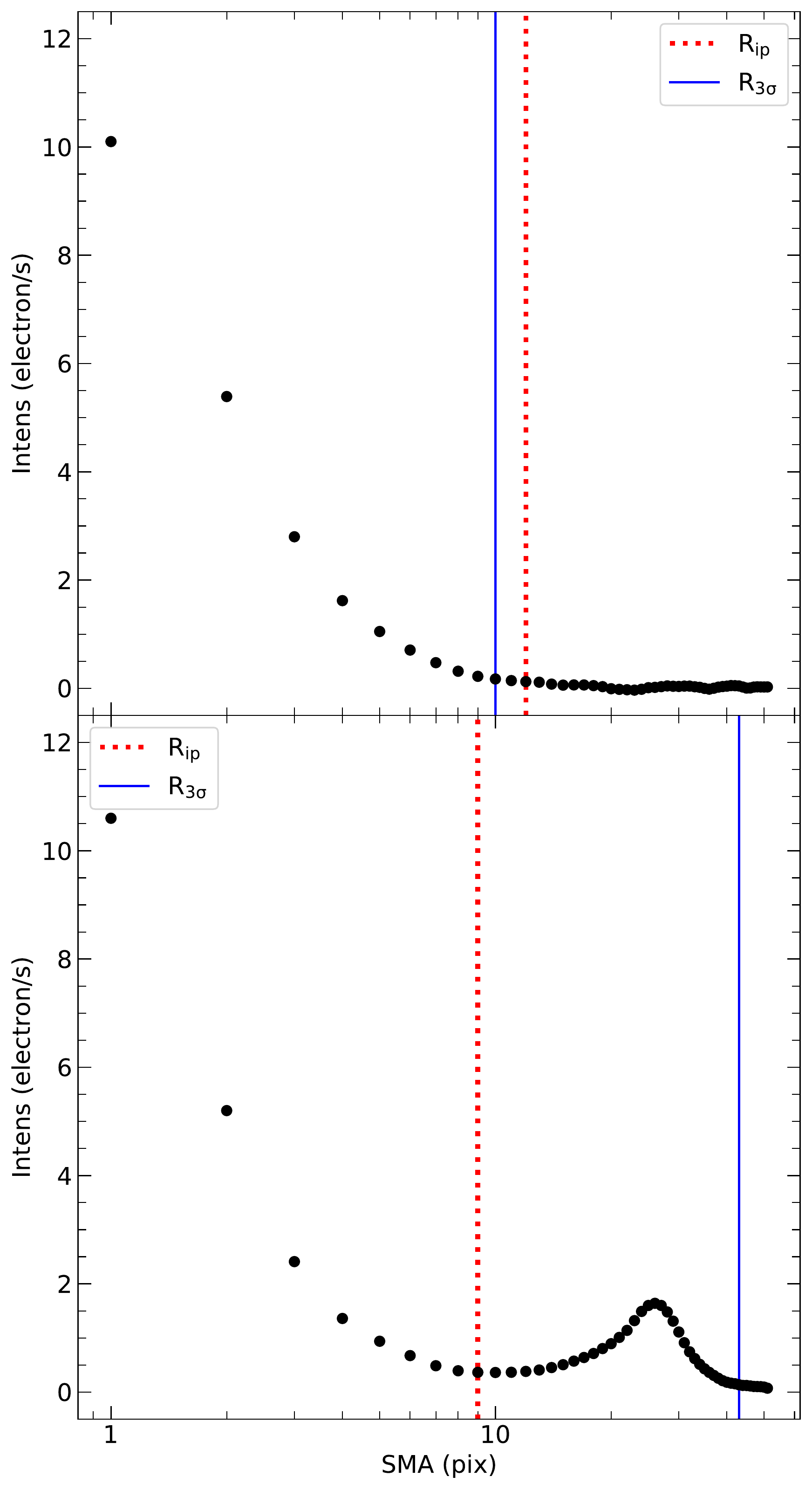}
\caption{Illustration of the choice of the fitting radius. The plots show isophotal intensity of two simulated clusters as a function of semi-major axes of the ellipses that best fit the isophotes. The top and bottom panels show clusters without and with a nearby contaminating source, respectively. The axis units correspond to natural units for the processed HST/ACS images, which is in electron/s for intensities and pixels of 0.05~arcsec for the semi-major axis (SMA).
}
\label{fig:rfit}
\end{figure}

\subsubsection{Fitting radius}

Considering that in the majority of datasets, clusters and extended objects in general are not isolated, it is necessary to remove the contribution of contaminant sources (even after masking them) in the fitting procedure. With this aim, we define the fitting radius. 
For isolated clusters, the background-subtracted SBPs are expected to monotonically decrease until the intensity values reach one $\sigma_{\rm bg}$, with  $\sigma_{\rm bg}$ being the dispersion of the measured background value intensity. We limit the fitting up to a radius at which the SBP has a value of $3\sigma_{\rm bg}$. We refer to such a radius as $R_{3\sigma}$. 
For objects located in relatively crowded regions, the SBPs show a bump instead of monotonically decreasing well before the intensity reaches the $3\sigma_{\rm bg}$ level. In such cases, we define $R_{\rm ip}$, the inflection point 
such that at $R_{ip}=\frac{d^2I}{dR^2}=0$. In Fig \ref{fig:rfit}, we illustrate the fitting radius selection in the case of a cluster with a contaminant source nearby as well as for an isolated cluster.  For each object under analysis, {\sc nProFit} computes both radii and sets as the fitting radius the minimum of $R_{3\sigma}$ and  $R_{\rm ip}$.

\subsection{Dynamical model fitting to observed SBPs}\label{Sec:modfit}

The main goal of our code is fitting the observed SBPs of star clusters to derive the basic structural parameters as well as dynamically relevant parameters. The structural parameters are obtained by fitting the observed SBPs with the theoretical SBPs for self-gravitating static models supported by pressure exerted by the stellar motions. We describe the theoretical SBPs implemented in {\sc nProFit} below.

\section{Theoretical SBPs}

The theoretical SBPs are related to the surface stellar density distributions (SDPs), through the mass-to-light ratio of stars. Clusters are self-gravitating and hence the SDPs determine the potential of the cluster. SDPs are obtained based on self-consistent potential-density pairs, given by phase-space distribution functions $f(x,v,t)$, depending on the positions and velocities of stars at a given time \citep{Galdynbook}.  Distribution functions (DF) allows us to derive the cluster's spatial density profile $\nu$   

\begin{equation}
\nu=\int f d^3v,
\end{equation}
and, the mean stellar velocity is defined by

\begin{equation}
\overline{v_i}=\frac{1}{\nu}\int f v_i d^3 v,
\end{equation}
$\nu$ is directly related to the luminosity profile through the mass-to-light ratio associated to  the objects' age \citep{BruzualCharlot2003}, and is constrained by the collisionless Boltzmann equation 

\begin{equation}
\frac{\partial f}{\partial t} + \sum_i a_i \frac{\partial f}{\partial v_i} + \sum_i v_i \frac{\partial f}{\partial x_i}=0,
\label{eq:boltzmann}
\end{equation}
with $x_i$, $v_i$ and $a_i$, the positions, velocities and accelerations of the stars.  It can be noticed that the Boltzmann equation depends on 7 variables, which hinders obtaining a direct solution of the equation. In order to simplify the problem, the Jeans Equation is used.

\begin{equation}
\frac{1}{\nu} \frac{d(\nu \overline{v_r^2})}{dr}+ 2 \frac{\beta\overline{v_r^2}}{r}=-\frac{d\phi}{dr},
\end{equation}
with $\beta$ the degree of anisotropy of the velocity distribution, $\overline{v_r^2}$ the radial velocity, $\phi$ the cluster potential. For the sake of this work, we assume isotropic velocity distributions, for which $\beta=0$, reducing the previous equation to two terms. The Jeans equation is obtained by taking moments of the Boltzmann Equation and integrating over all the velocities.

\subsection{King models}

King models are based on a modified isothermal sphere. The density of the dynamical King models is given by a distribution function \citep{King_dyn}.  The King distribution is described as follows 

\begin{equation}
f(\varepsilon) \propto \left\{ 
\begin{array}{cc}
\exp{(\varepsilon/\sigma_0^2)} -1, & \varepsilon>0,\\
0, & \varepsilon \leq 0,   
\end{array}
\right. ,
\label{distfuncking}
\end{equation}
with $\varepsilon$ the relative energy of the system (defined in terms of the central gravitational potential and the total system energy as $\varepsilon=-E+\phi_0$, (or equivalently $\varepsilon=\Psi -\frac{1}{2}v^2$) 
per unit mass, $\sigma_o$ the model-associated velocity dispersion described as follows

\begin{equation}
\sigma_0^2 \equiv \frac{4\pi G \rho_0 r_0^2}{9},
\label{relkingsigma}
\end{equation}
with $\rho_0$ the central density and $r_0$ the scale factor, referred to as the King Radius in the literature. 

\begin{figure*}
\centering
\includegraphics[width=\textwidth]{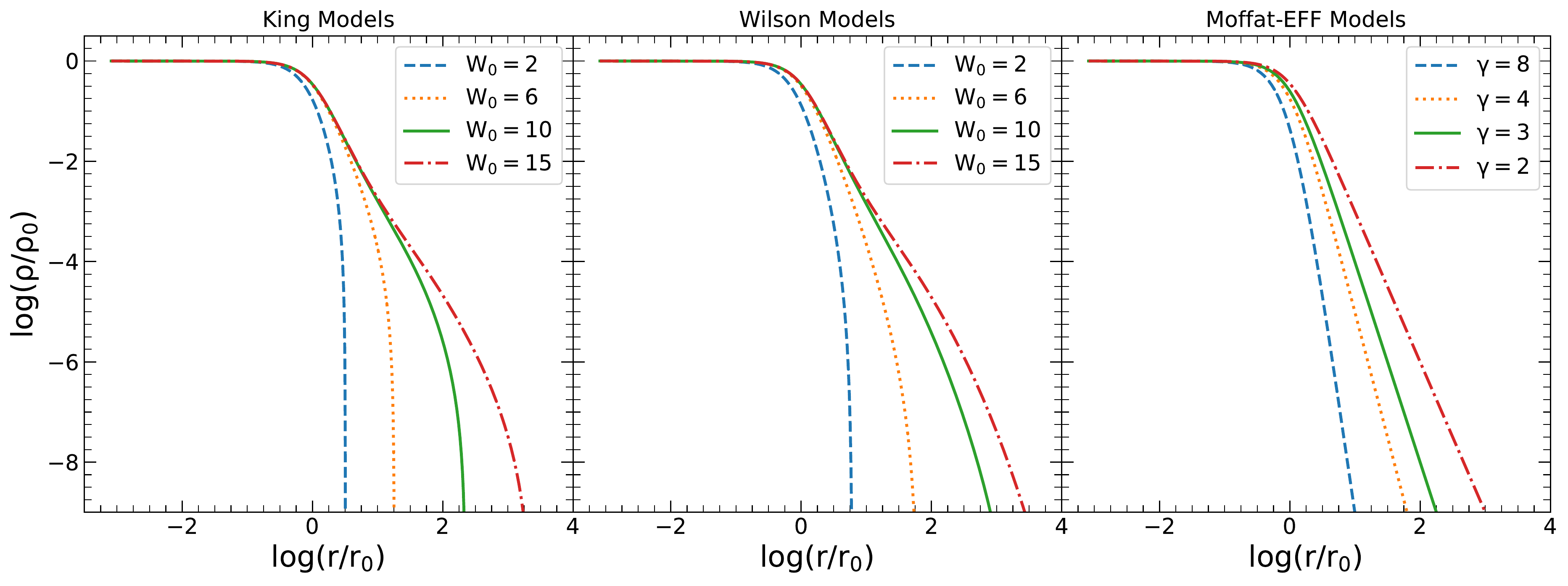}
\caption{Normalized volume densities corresponding to King (left panel), Wilson (middle panel) and Moffat-EFF (right panel).  The densities corresponding to isothermal-based models (King and Wilson), were computed by {\sc nProFit}.}
\label{fig:models}
\end{figure*}

From the Poisson's equation (Eq. \ref{eq:pois_boltz}) 

\begin{equation}
\nabla^2 \phi= 4\pi G \rho= 4\pi G \int fd^3 v.
\label{eq:pois_boltz}
\end{equation}
and Eq. \ref{distfuncking}, setting $W=\frac{\psi}{\sigma^2}$, the density function is obtained, in terms of $W$, the dimensionless potential

\begin{equation}
\rho(W)= \frac{15}{2} e^W erf (\sqrt{W})\sqrt{\pi} -\sqrt{W}\bigg{(}W+\frac{3}{2}\bigg{)},
\label{rho_king_eq}
\end{equation}
with erf(x) the error function $erf(x)=\frac{2}{\sqrt{\pi}} \int_o^x e^{-t^2}dt$.  $W$ is obtained by solving $\nabla^2 W=4\pi G\rho(W)$, with boundary conditions $W(0)=W_0$, with $W_0$ the central potential, and $W(r_t)=0$, with $\it r_t$ the tidal radius.  

In Fig. \ref{fig:models} (left panel), we show the density profiles for 4 King models, obtained from the previously described procedure.  We show concentrated ($W_0=2$ and $W_0=6$) as well as more extended profiles ($W_0=10$ and $W_0=15$).

\subsection{Wilson models}

Wilson models are based on  isothermal sphere models as the King models.  These models were originally proposed by \citet{Wilson_dyn} to fit the observed surface brightness profiles of elliptical galaxies, having larger haloes than the King models, produced by an extra term in the energy distribution function.  The distribution function of these models is as follows
\begin{equation}
f(\varepsilon) \propto \left\{ 
\begin{array}{cc}
\exp{(\varepsilon/\sigma_0^2)} -1-\frac{\varepsilon}{\sigma_0^2}, & \varepsilon>0,\\
0, & \varepsilon\leq 0,   
\end{array}
\right. ,
\label{distfuncwilson}
\end{equation} 
with $\varepsilon$ the relative energy of the system, and $\sigma_o$ the model-associated velocity dispersion described in Eq. \ref{relkingsigma}.\\

The corresponding mass density function is obtained proceeding analogously as in the King model, and follows

\begin{equation}
\rho(W)=e^W erf (\sqrt{W}) - \bigg{(}2\sqrt{\frac{W}{\pi}}\bigg{)} \bigg{(} \frac{4W^2}{15}+\frac{2W}{3}+1\bigg{)},
\label{rho_wilson_eq}
\end{equation}

The theoretical surface brightness for King and Wilson models can be found from the mass density function by means of the following integral

\begin{equation}
I(R) = \frac{\Sigma(R)}{\zeta} = \frac{2}{\zeta} \int_R ^{R_t} \frac{\rho(r)}{(r^2-R^2)^\frac{1}{2}} rdr,
\label{rho_surf_king_eq}
\end{equation}
with $\zeta$ the mass-to-light ratio. For the sake of simplicity, we assume $\zeta=\rm 1~M_\odot/L_\odot$ throughout this work.

In Fig. \ref{fig:models} (middle panel), we show the density profiles for 4 Wilson models, obtained from the previously described procedure.  We show concentrated ($W_0=2$ and $W_0=6$) as well as more extended profiles ($W_0=10$ and $W_0=15$).

\subsection{Moffat-EFF empirical profiles}

In the pioneering work by \citet{Elson}, King models were used to fit the observed profiles of 10 intermediate-age star clusters in the Large Magellanic Cloud, that have masses and densities similar to that of old Globular Clusters \citep{Portegiesrev}. They found King models did not provide good fits to the SBPs because these clusters displayed more extended haloes instead of truncated outer parts.  \citet{Elson} proposed an empirical profile, based on Moffat \citep{Moffat1969} profile (henceforth Moffat-EFF where EFF stands for Elson, Freeman and Fall), given by

\begin{equation}
I(R)=\frac{(\gamma-2) L_{\rm tot}}{2 \pi r_{\rm d}^2} \bigg{[}1+\bigg{(}\frac{R}{r_{\rm d}}\bigg{)}^2\bigg{]}^{-\gamma/2}, 
\label{moffat_eq}
\end{equation}
with $R$ the observed profile projected semi-major axis, $r_d$ the characteristic radius or model scale radius,  $L_{tot}$ the total luminosity, and $\gamma$ the Moffat-EFF index, which provides information on the shape of the halo.

Moffat-EFF profile does not have an implicit distribution function. However, its 3-D luminosity density profile can be calculated using the expression

\begin{equation}
j(r)= j_{\rm 0} \bigg{(}1+\frac{r^2}{r_{\rm d}^2}\bigg{)}^{-(\gamma+1)/2},
\end{equation}

\subsection{Volume density and dispersion velocity profiles for isothermal spheres}\label{subsec:theo_sbs}

The King and Wilson models compute the $\rho(r)$ as a function of a numerically defined potential $W$ for isothermal spheres, given by equation \ref{rho_king_eq} and \ref{rho_wilson_eq}, respectively. The $\rho(r)$ and $W$ define the core radius, dynamical mass, surface and volume mass densities, binding energy, bound mass and central velocity dispersion, and hence the dynamically useful parameters corresponding to the best-fit model are known a priori. 

In order to obtain $\rho(W)$, it is necessary to solve the following expression, obtained from the Laplacian operator in spherical coordinates and in terms of the dimensionless potential $W$ 

\begin{equation}
\nabla^2 \phi=  \frac{1}{r} \frac{\partial}{\partial r} \bigg{(}r \frac{\partial W}{\partial r}\bigg{)},
\end{equation}
and since $W$ depends only on $r$, and following the prescription by \citet{King_dyn} the Poisson's equation follows
\begin{equation}
\frac{d^2 W}{d r'^2}+\frac{2}{r'}\frac{dW}{dr'}= -9 \frac{\rho}{\rho_0}, 
\end{equation}
with $r'=r/r_0$.   
With the aim of solving these differential equations, we proceed to re-write the equation with $W$ becoming the independent variable, to be consistent with  Eqs. \ref{rho_king_eq} and \ref{rho_wilson_eq}, following \citet{King_dyn,HeggieAarseth1992,Mcluster} 
\begin{equation}
-X\frac{d^2 X}{d W^2}+\frac{3}{2}\bigg{(}\frac{dX}{dW}\bigg{)}^2= -\frac{9}{4} \frac{\rho}{\rho_0} \bigg{(}\frac{dX}{dW}\bigg{)}^3,
\label{eq:king_k4_a}
\end{equation}
with $X=r'^2$.  Following the dependence of $\rho$ on $W$, the first step is to use the 4th order Runge-Kutta method \citep{runge1895,kutta1901} to obtain $W$ in terms of $X$ and, thus $\rho$ in terms of $W$ for each obtained value from each iteration, following
\begin{equation}
y_{n+1}=y_n+\frac{h}{6}(k_1+2k_2+2k_3+k_4),
\end{equation} 
where,
\begin{equation}
\begin{matrix}
k_1=f(x_n,y_n),\\
k_2= f(x_n+\frac{h}{2},y_n+\frac{k_1}{2}),\\
k_3= f(x_n+\frac{h}{2},y_n+\frac{k_2}{2}),\\
k_4= f(x_n+h,y_n+k_3).\\
\end{matrix}
\label{eq:ks_runge}
\end{equation}  
Re-writing Eq. \ref{eq:king_k4_a} as
\begin{equation}
\frac{d^2 X}{d W^2}=\frac{1}{4X}\bigg{(}\frac{dX}{dW}\bigg{)}^2 \bigg{[}6+9 \frac{dX}{dW}\frac{\rho}{\rho_0} \bigg{]},
\label{eq:king_k4}
\end{equation}
and setting $y=X$, we have $y'=\frac{dX}{dW}$, and $y''=\frac{d^2X}{dW^2}$. The latter expression can be expressed by substituting Eq. \ref{eq:king_k4} 
\begin{equation}
y''=\frac{1}{4X}\bigg{(}\frac{dX}{dW}\bigg{)}^2 \bigg{[}6+9 \frac{dX}{dW}\frac{\rho}{\rho_0} \bigg{]}.
\end{equation}
Finally we proceed to apply Runge-Kutta two times to obtain the values of $W$ 
\begin{equation}
W_{n+1}=W_n+\frac{h}{6}(k_1+2k_2+2k_3+k_4),
\end{equation}
and
\begin{equation}
r'_{n+1}=r'_n+\frac{h}{6}(k_1'+2k_2'+2k_3'+k_4'),
\end{equation} 
with $k_1$, $k_2$, $k_3$ and $k_4$, and $k_1'$, $k_2'$, $k_3'$ and $k_4'$ following Eq. \ref{eq:ks_runge}, with $f=y'$ and $f=y''$ respectively.  From the values of $W$, the mass volume density of King and Wilson models are obtained following Eqs. \ref{rho_king_eq} and \ref{rho_wilson_eq}, respectively.

For King and Wilson models, {\sc nProFit} calculates the corresponding profiles following the prescription in \citet{Galdynbook} where
\begin{equation}
\overline{v^2(r)}=\frac{J_2}{J_0}, \quad {\rm with} \quad  J_n \equiv \int_0 ^{\sqrt{2 \varepsilon \sigma^2}} f(\varepsilon) v^{n+2} dv.
\end{equation}
resulting in the following equations for King and Wilson profiles, respectively

\begin{equation}
\overline{v^2(r)}=\frac{6\sigma_0^2}{5}  \frac{-\sqrt{W} (W^2+\frac{5}{2}W+ \frac{15}{8}) + \frac{15\sqrt{\pi}}{4} e^{W}  erf(\sqrt{W})}{-\sqrt{W} (W+ \frac{3}{2}) + \frac{3\sqrt{\pi}}{4} e^{W}  erf(\sqrt{W})} 
\end{equation}

\begin{equation}
\resizebox{\hsize}{!}{$\overline{v^2(r)}=\frac{6\sigma_0^2}{7} \frac{-\sqrt{W} (W^3+\frac{7}{2} W^2-\frac{35}{4}W+\frac{105}{8})+\frac{105\sqrt{\pi}}{16} e^W  erf(\sqrt{W})}{-\sqrt{W}(W^2+\frac{5}{2}W+\frac{15}{4}) + \frac{15\sqrt{\pi}}{8} e^W erf(\sqrt{W})}$}
\end{equation}

\begin{figure}
\centering
\includegraphics[width=0.8\columnwidth]{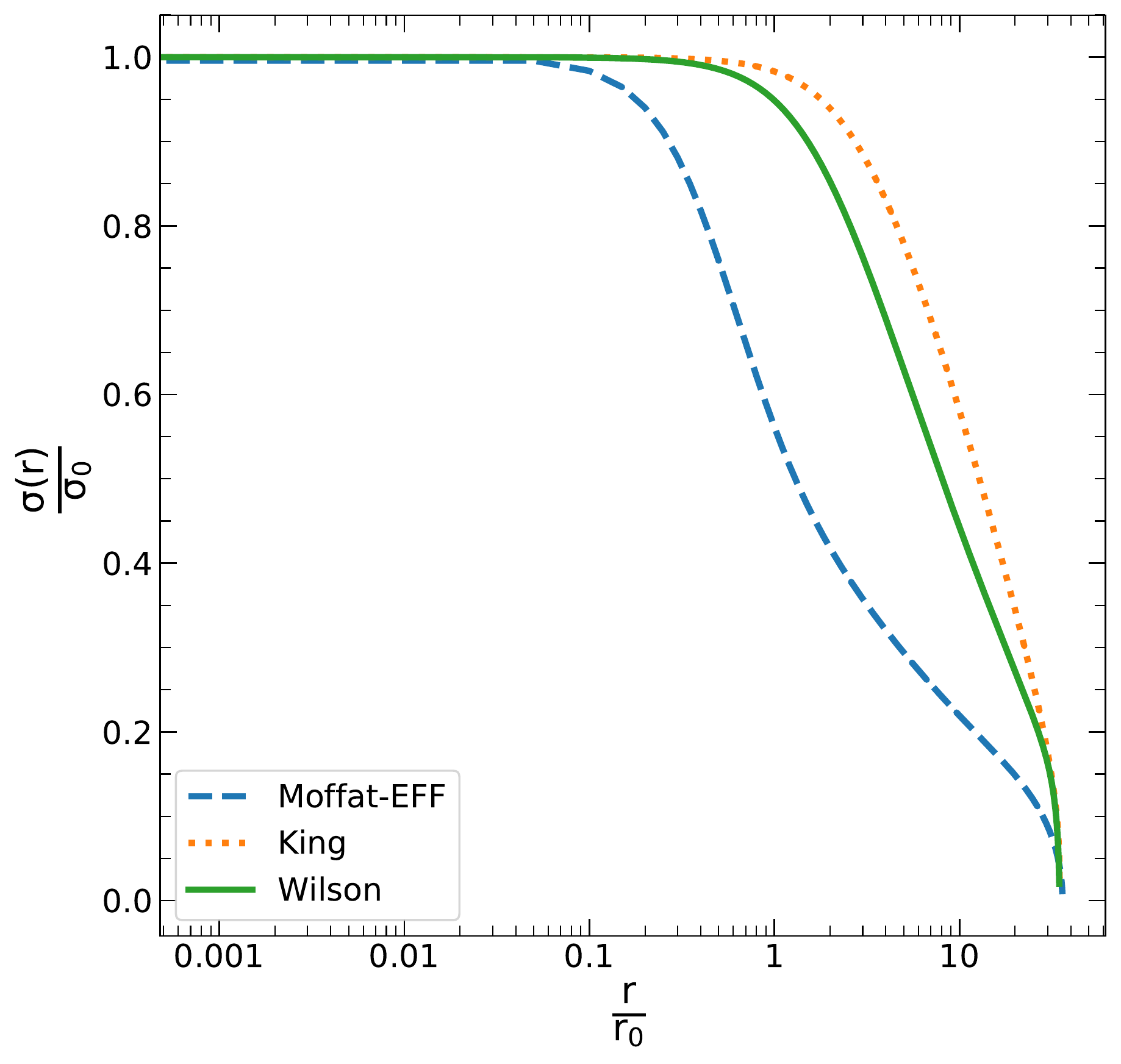}
\caption{Velocity dispersion profiles for King, Wilson and Moffat-EFF models, with the same tidal radius $r_t$, and volume density $\rho=10^{4.2}$ $\rm M_\odot/pc^3$. The King and Wilson profiles have $W_0=7.1$ and $W_0=5.7$, respectively, and the corresponding $\gamma$ index for the Moffat-EFF profile is 2.7}
\label{fig:disp_profs}
\end{figure}

In Fig. \ref{fig:disp_profs}, we show the velocity dispersion profiles for King and Wilson models with $W_0=7.1$ and $W_0=5.7$, with the same tidal radius $r_t$ for illustration purposes.

\subsection{Potential and velocity dispersion profiles for Moffat-EFF models}{\label{sec:sig_moff}}
   
On the other hand, some initially empirical models, such as the Moffat-EFF model, can be physically motivated.  For instance, by means of Eq. \ref{moffat_eq}, the Moffat-EFF volume density yields 

\begin{equation}
\rho(r)=\rho_{\rm 0} \bigg{(}1+\frac{r^2}{r_{\rm d}^2}\bigg{)}^{-(\gamma+1)/2},
\label{moffat_eq2}
\end{equation}
where 

\begin{equation}
\rho_{\rm 0}=\frac{I_0\, \Gamma((\gamma+1)/2) \zeta }{\sqrt{\pi}\,  \Gamma(\gamma/2)\, r_{\rm d}},
\label{eq:dens:moff}
\end{equation}
with, $I_0$ the central surface brightness in units of $L_\odot$/pc$^2$ and $\Gamma$ the usual gamma function. We bear in mind that in general, mass profiles do not strictly follow light profiles over all radii due to the effects of mass segregation \citep{ShanahanGieles,Baumgardt2017}, resulting in mass functions flatter than the luminosity functions. However, for the sake of simplicity, we assume that mass profiles follow light profiles, and convert from such quantities by means of the mass-to-light ratios.

In Fig. \ref{fig:models} (right panel), we show the density profiles for 4 Moffat-EFF models, obtained from the previously described procedure.  We show concentrated ($\gamma=4$ and $\gamma=8$) as well as more extended profiles ($\gamma=2$ and $\gamma=3$).

{\sc nProFit} also computes the theoretical velocity dispersion profile for Moffat-EFF model following the Eq. 16 in the prescription by \citet{Elson}, under the assumption of a spherical cluster under hydrostatic equilibrium
\begin{equation}
\sigma^2(r) = \frac{G}{r_d} \bigg{(} 1+ \frac{r^2}{r_d^2} \bigg{)}^{\frac{(1+\gamma)}{2}} g  \bigg{(} \frac{r}{r_d}  \bigg{)} - \frac{r_d^2(4 \Omega^2-\kappa^2)}{(\gamma-1)}  \bigg{(} 1 + \frac{r^2}{r_d^2} \bigg{)}
\end{equation}
with 
\begin{equation}
g\bigg{(}\frac{x}{r_d}\bigg{)}=\int_{\frac{x}{r_d}}^\infty \frac{M(x') dx'}{\big{(}\frac{x'}{r_d}\big{)}^2 \big{(}1+(\frac{x'}{r_d})^2 \big{)}^\frac{(\gamma+1)}{2}}
\end{equation}
and $\Omega$ and $\kappa$ the circular and epicyclic frequencies of the galaxy at the pericenter of the orbit of the cluster, and $M(x)$ the mass enclosed at radius $x$.

In Fig. \ref{fig:disp_profs}, we show the velocity dispersion profile for a Moffat-EFF profile with $\gamma=2.7$, compared with King and Wilson profiles with the same $r_t$.

The corresponding potential $\phi$ for Moffat-EFF model is computed  by means of the Poisson Equation in spherical coordinates resulting in the following equation

\begin{equation}
\begin{array}{l}
\phi(r)= 4 \sqrt{\pi}  G\, \zeta  r_d\, \mu_0 \bigg{(}1+\frac{r^2}{r_d^2}\bigg{)}^{\frac{(1-\gamma)}{2}}  \bigg{(}1+\frac{r_d^2}{r^2}\bigg{)}^{\frac{(\gamma-1)}{2}}  f_\Gamma(r)\\
\\
{\rm where}\\
\quad \quad \quad \quad f_\Gamma(r)=\frac{\Gamma(\frac{\beta+1}{2}){}_2F_1\Big{[}\frac{\gamma-1}{2},\frac{\gamma-1}{2},\frac{\gamma+1}{2},-\frac{r_d^2}{r^2}\Big{]}}{(\gamma-1)^2\Gamma(\frac{1}{2} \gamma)}
\end{array}
\label{eq:poten}
\end{equation} 
with ${}_2F_1$ the hypergeometric function and $\Gamma$ the usual gamma function.

\subsection{Derived parameters}\label{Sec:der_pars}

We now use our results for best-fit models to extract the most commonly used structural parameter, namely core radius and half-light radius. The latter quantity depends on the concentration parameter for the King and Wilson models, and on the gamma for the Moffat-EFF profile. We also provide three additional parameters namely projected central velocity dispersion, total mass and binding energy of the clusters, for an assumed mass-to-light ratio of 1.0

\subsubsection{Concentration index $c$} 

\begin{figure}
\centering
\includegraphics[width=0.8\columnwidth]{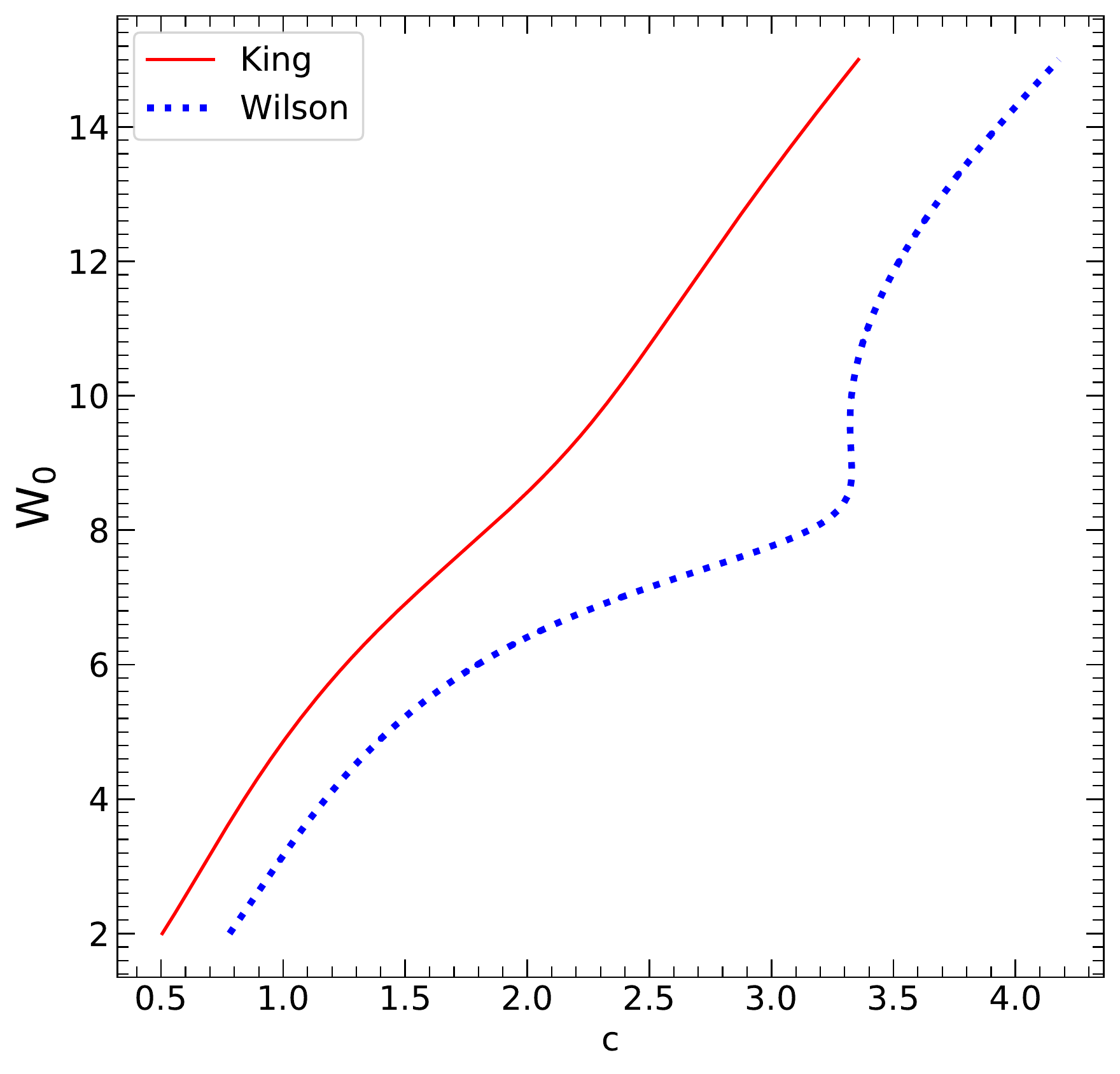}
\caption{Relation between the concentration index $c$ and the dimensionless potential $W_0$, computed by {\sc nProFit} following the procedure described in Sec. \ref{subsec:theo_sbs} for King and Wilson models.}
\label{fig:W0_vs_c}
\end{figure}

For \citet{King_dyn} and \citet{Wilson_dyn} models, the central potential $W_0$ is related to the concentration parameter $c=\log(r_t/r_0)$ obtained by fitting empirical \citet{King_emp} formula. In Figure \ref{fig:W0_vs_c}, we show this relation, which is obtained from the solutions of the models described in the previous section.

\begin{figure*}
\centering
\includegraphics[width=\textwidth]{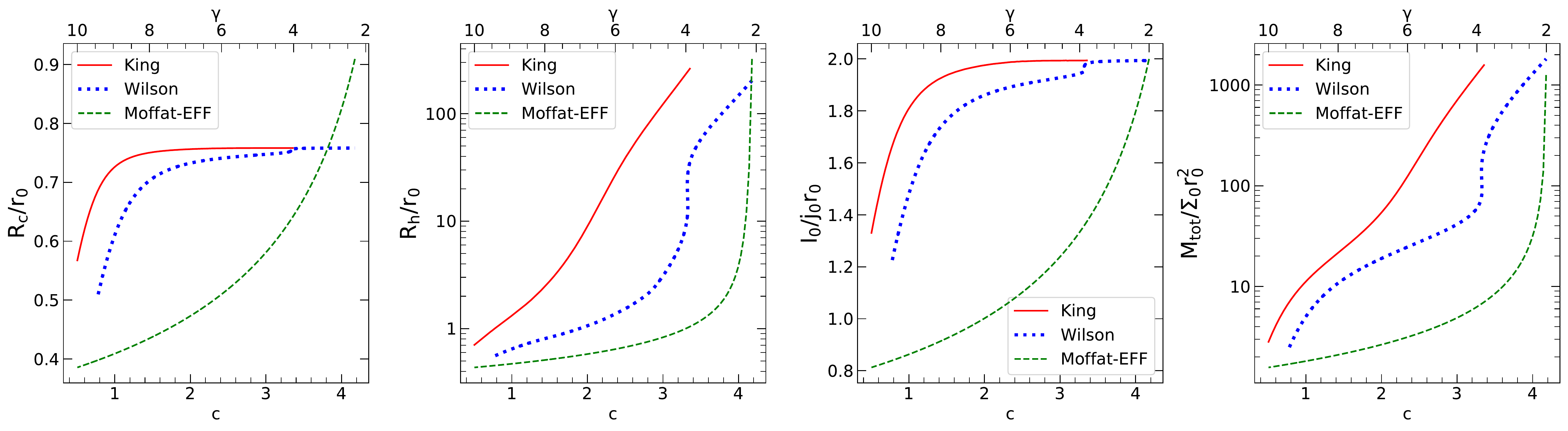}
\caption{Normalised functions used to compute $\rm R_c$ (left-most panel), $\rm R_h$ (second panel from left to right), and auxiliar functions for $\rm I_0/j_o r_0$ (third panel) and $\rm M_{tot}$ (right-most panel).  These functions are used by {\sc nProFit}, to speed up calculations.}
\label{fig:funcs_aux}
\end{figure*}

\subsubsection{Core radius $R_c$}

The scale size of the isothermal spheres ($r_0$), and Moffat-EFF profiles ($r_d$) are related to the core radius $R_c$. This quantity is computed recalling that $R_c$ is the radius at which the luminosity density reaches half its peak value.

For Moffat-EFF profiles, $R_c$ is given in terms of the characteristic radius or model scale radius ($r_d$) as follows

\begin{equation}
R_c=r_d (2^{2/\gamma}-1)^{1/2}.
\label{rc_rd}
\end{equation}

The corresponding values for King and Wilson models are obtained by {\sc nProFit} by interpolating over the profile density values.  In the left-most panel of Fig. \ref{fig:funcs_aux}, we show an auxiliar dimensionless value used by {\sc nProFit} to compute $R_{\rm c}$  for King and Wilson models. For the sake of completeness, we show the same dimensionless function in terms of $\gamma$ for Moffat-EFF models.

\subsubsection{Half-light radius $R_h$}{\label{sec:rh}}

For a given core size of isothermal spheres, the half-mass radius is related to the concentration index. We show such a relation in the second panel from left to right in Fig. \ref{fig:funcs_aux}, with Wilson models having larger values at the same $R_h$ value due to the more extended profiles of the Wilson models.  \citet{McLaughlin2000} characterized such a relation by fitting a 9th order polynomial for King models.   We have also performed a polynomial fitting, in this case with a 5th order polynomial, for both King (Eq. \ref{eq:King_c_vs_rh}) and Wilson (Eqs. \ref{eq:Wilson1_c_vs_rh} and \ref{eq:Wilson2_c_vs_rh}, for values below or equal and above $c=3.25$, respectively) models.

\begin{equation}
\begin{split}
\log\bigg{(}\frac{R_h}{r_0}\bigg{)}=0.07105 c^5 -0.7433 c^4+2.814 c^3\\ 
-4.552 c^2+3.754c -1.226,
\end{split}
\label{eq:King_c_vs_rh}
\end{equation}
if $c\leq 3.25$
\begin{equation}
\begin{split}
\log\bigg{(}\frac{R_h}{r_0}\bigg{)}= 0.05389 c^5-0.4788 c^4+1.708 c^3\\
-3.007 c^2+2.79 c-1.253,
\end{split}
\label{eq:Wilson1_c_vs_rh}
\end{equation}
otherwise
\begin{equation}
\begin{split}
\log\bigg{(}\frac{R_h}{r_0}\bigg{)}=50.3330062 c^5-953.958025 c^4\\
+7223.17915 c^3-27312.3861 c^2\\
+51573.8755 c-38906.3596. 
\end{split}
\label{eq:Wilson2_c_vs_rh}
\end{equation}

On the other hand, for the Moffat-EFF profile, the $R_h$ is analytically related to the fitted structural parameters $r_{\rm d}$ and $\gamma$.

\begin{equation}
R_h = r_{\rm d}(0.5^{1/(1-\gamma/2)}-1)^{1/2}.
\label{eq:Rh_cuevas}
\end{equation}

We show the dimensionless function $R_h/r_0$ for illustration purposes in the second panel from left to right in Fig. \ref{fig:funcs_aux}, along with the corresponding $R_h/r_d$ relatiion for Moffat-EFF models, for the sake of completeness.

\subsubsection{Central velocity dispersion $\sigma_0$}

{\sc nProFit} computes de central velocity dispersion values from the previously computed velocity profiles described in Sec. \ref{subsec:theo_sbs} for King and Wilson models and \ref{sec:sig_moff} for Moffat-EFF empirical profiles.  The central velocity dispersion profile projected into the plane of the sky $\sigma_{p,0}$ is computed for isothermal models as well as for Moffat-EFF models following 

\begin{equation}
\sigma_{p}(R)=\frac{2}{I(R)}  \int_R^\infty \frac{j \overline{v_r^2} r dr}{\sqrt{r^2-R^2}},
\end{equation}
with $\overline{v_r^2}$ the quadratic velocity dispersion profile computed in Sec. \ref{subsec:theo_sbs} and \ref{sec:sig_moff}, and $I(R)$ the SBP. 

\subsubsection{Tidal Radius $\rm r_t$}

From the computed velocity dispersion profiles corresponding to Moffat-EFF models, we can compute the tidal radius, by finding the radius $r$ at which $\sigma_p(r)=0$. On the other, hand, we find such quantity for King and Wilson models, from the concentration parameter and scale radius, following the expression for the concentration index $r_t=10^c\, r_0$.

\subsubsection{Total mass $M_{tot}$}

Another critical parameter these models provide is the total mass.  For Moffat-EFF models, {\sc nProFit} derives the model mass from the total luminosity $L_{tot}$, using Eq. \ref{moffat_eq}  and assuming a mass-to-light ratio $\zeta$.  The mass corresponding to King and Wilson models is computed from Eqs. 38 and 40 in \citet{King_dyn} instead, resulting in 

\begin{equation}
M_{tot}= 4 \pi \zeta j_0 r_0^2 \int_0^{r_t'} j' r'^2 dr'    
\end{equation}
with $j'=\frac{j}{j_0}$ (expressed as a normalised quantity, as obtained in the previous sections), $r_t'=\frac{r_t}{r_0}$ and $r'=\frac{r}{r_0}$. In the right-most panel in Fig.  \ref{fig:funcs_aux}, we show auxiliar dimensionless function used by {\sc nProFit} to compute $M_{tot}$ for King and Wilson models, along with Moffat-EFF profiles for completeness..

\begin{figure}
\centering
\includegraphics[width=0.8\columnwidth]{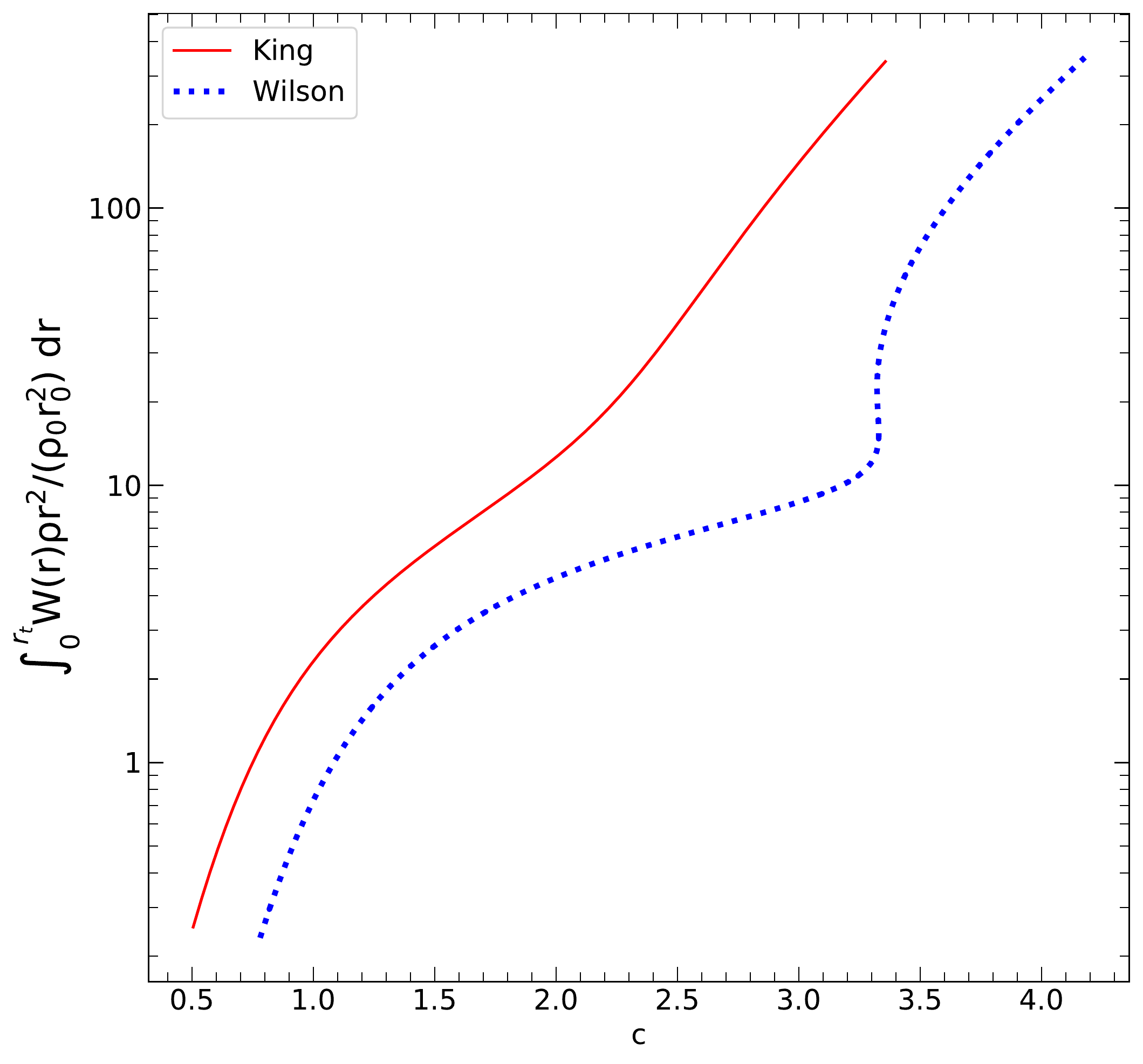}
\caption{Auxiliar function to compute the integral in Eq. \ref{eq:ebin_k} for the binding energy.}
\label{fig:ebin_aux}
\end{figure}

\subsubsection{Binding energy $E_b$}

Following the prescription by \citet{McLaughlin2000}, we have also implemented in {\sc nProFit} the computation of the binding energy, ($\rm E_b$) using their Eq. 1  

{\sc nProFit} computes $\rm E_b$ for Moffat-EFF models, substituting the potential described in Eq. \ref{eq:poten} along with the density $\rho$ in Eq. \ref{moffat_eq2} into the following equation 

\begin{equation}
E_b= - \frac{1}{2} \int_0^{r_t} 4\pi \rho \phi dr    
\label{eq:ebin_m}
\end{equation}

On the other hand, for King and Wilson models,  {\sc nProFit} uses the dimensionless function in the right-most panel in Fig. \ref{fig:ebin_aux} to compute the integral in the following equation 
\begin{equation}
E_b=\frac{1}{2} \int_0^{r_t} 4 \pi r^2 \rho \bigg{[} \frac{GM}{r_t} + \sigma_0^2 W(r) \bigg{]} dr    
\label{eq:ebin_k}
\end{equation}

\subsubsection{Central surface magnitude $\mu_0$}

The central surface brightness $I_0$ in $L_\odot$\,pc$^{-2}$ is converted to observational units ($\rm \mu_{F,0}$) using

\begin{equation} 
\rm \mu_{F,0}=M_{F,\odot} +21.57 -2.5 \log{I_0}
\end{equation}
with $M_{F,\odot}$ the absolute magnitude of the Sun in the filter F, provided by the user (see \citet{Christopher2018}).

\subsection{Library of dynamical models}

For King and Wilson models, {\sc nProFit} solves the previously described set of differential equations with boundary conditions for $W_0\in$[2,15] in steps of 0.1 in terms of $\frac{r}{r_0}$. The latter results in a library constituted by files containing  $\frac{r}{r_0}$, $W$, $\rho$, and $\Sigma$. These libraries are the results of the most computer-intensive module of {\sc nProFit}.  They are pre-evaluated to speed up computation times and the corresponding results are used to compute the previously described derived parameters. {\sc nProFit} uses this library for any new fit, unless the user specifically asks to solve the equations.

\begin{figure}[ht]
\includegraphics[width=\columnwidth]{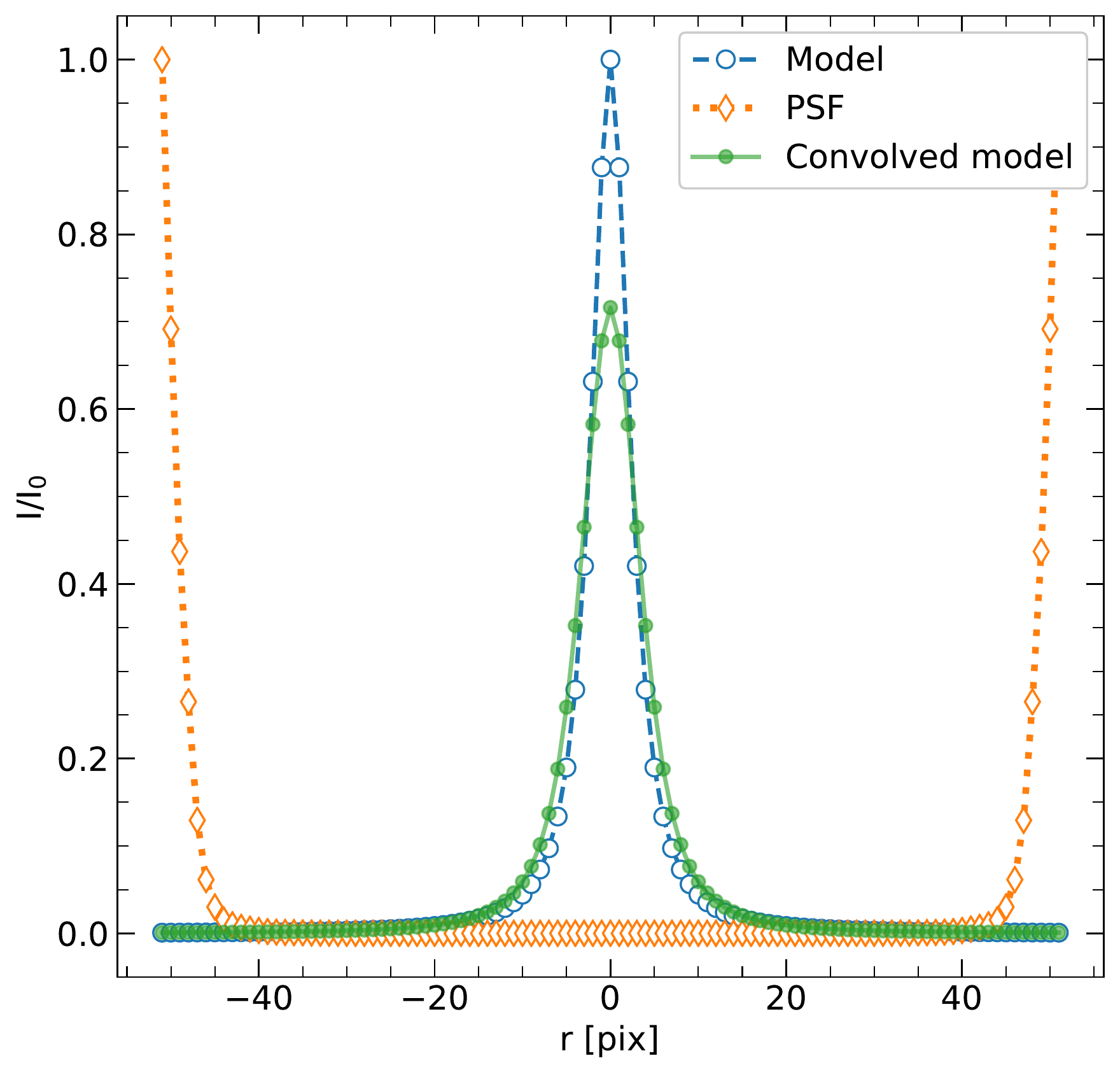}
\caption{Convolution scheme performed by {\sc nProFit}. The model is wrapped around and the PSF is mirrored and zero-padding is performed to meet the required size criterion for the FFT. The plotted example corresponds to $2^7=128$ pixels.} 
\label{fig:conv}
\end{figure}

\subsection{Convolution with the Point Spread Function (PSF)}

The theoretical profiles are convolved with the user-provided PSF\footnote{PSF for the HST images can be obtained using the Tinytim tool (see https://www.stsci.edu/hst/instrumentation/focus-and-pointing/focus/tiny-tim-hst-psf-modeling). Alternatively, it can be defined for each frame using the tasks for that purpose in {\sc daophot} package or Sextractor command PSFex. In the present work, we assumed a Gaussian profile of FWHM=2.1 pixels, which represents well the point sources on the HST/ACS images.}
by {\sc nProFit} using a {\sc Fortran} routine.  We recall that in general, for finite datasets as the models we described, the convolution operator is given by

\begin{equation}
(g*h)(m) =\sum_{n=-N}^N g(n) h(m- n),
\label{eq:conv_disc}
\end{equation}
with $g$ the signal and $f$ the response function, which in our case are the dynamical models and the PSF, respectively.  Computing  Eq. \ref{eq:conv_disc} requires $N^2$ operations to perform the convolution.  In order to reduce the number of operations, {\sc nProFit} uses the fast fourier transform (FFT), resulting in $N\log_2 N$ operations, reducing the number of operations in 65\%.  {\sc nProFit} computes the FFT using the numerical recipes routine {\sc convlv} \citep{Numrec}.  In general, the FFT splits the convolution expression in terms of odd and even indices in the sum.  Hence,  FFT routines require the input sampled in number of points being a power of 2.  It is also required for the {\sc convlv} routine, 
that the PSF or response function is wrapped around and filled with zeros to obtain a vector with a length of N-1 (odd size). N is the size of the vector containing the dynamical model, which is a power of two.  The model needs to be mirrored and wrapped around. In Fig \ref{fig:conv}, we illustrate the convolution procedure along with the required zero-padding and wrapping.

\subsection{Selection of the best-fit model}

From the previously determined fitting radius, {\sc nProFit} proceeds to compute which model provides the most accurate and reliable representation of the observed SBPs.  To this aim, we use the non-parametric statistical test $\chi^2$ \citep{Bevington1993}.  
The $\chi^2$ test determines the goodness of a fit to data, suitable for observed data with gaussian errors, which by virtue of the central limit theorem \citep{Mood_estadistica} are good approximations of Poisson distributions.  Poisson statistics are crucial in the determination of the data noise, having different variances, requiring to use a $\chi^2$ test  weighted by errors \citep{Wall_statistics}
\begin{equation}
\chi^2=\sum_{i=1}^{N{\rm pts}} \frac{(I_{{\rm obs}_i}-\tilde{I}_{{\rm model}_i})^2}{\sigma_i^2},
\label{eq:chi_sq}
\end{equation}

\noindent with $N{\rm pts}$ the number of points, the azimuthally-averaged profile intensities $I_{\rm obs_i}$, $\sigma_i$ the corresponding errors computed during the isophotal fitting, and $\tilde{I}_{\rm model_i}$ the PSF-convolved model intensities at $i$, varying from 1 to  $N{\rm pts}$.
{\sc nProFit} performs the $\chi^2$ minimization technique to find the best-fit model in the parameter space for each one of the available theoretical models in {\sc nProFit}. In order to determine which of these models represents most accurately the observed data, we implemented the prescription by \citet{McLaughlin2005} to compare the obtained best-fit models 

\begin{equation}
\Delta \chi^2 = \frac{\chi^2_{\rm alt} - \chi^2_{\rm ref}}{\chi^2_{\rm alt} + \chi^2_{\rm ref}},
\label{eq:delta_chi}
\end{equation}
where $\chi^2_{\rm ref}$ and $\chi^2_{\rm alt}$ are the $\chi^2_{\rm min}$ values of the reference model and the model to be compared, respectively.

This procedure allows to determine in which cases two models are equally good $|\Delta\chi^2|\leq$0.2 or whether the alternative model provides a better fit than the reference model.

\subsection{Determination of errors}

{\sc nProFit} computes the errors on the obtained parameters by considering 
1-$\sigma$ significance regions. Considering that each fit depends basically on three free parameters, the intervals for a $1-\sigma$ level of confidence are defined by means of the following equation \citep{Wall_statistics}

\begin{equation}
\chi^2=\chi^2_{min}+3.50,
\end{equation}
with  $\chi^2_{min}$ the $\chi^2$ value of the best fit model, and $\chi^2$ the value corresponding to the rest of the models. These intervals are computed for Moffat-EFF, King and Wilson models, separately.  The errors on the derived parameters described in Sec. \ref{Sec:der_pars} are computed by {\sc nProFit} by propagating the errors on the basic parameters for each models, namely, $r_d$ and $\gamma$ for Moffat-EFF models and $r_0$ and $W_0$ for King and Wilson models. 

\section{Code illustration with simulated clusters}{\label{Sec:nprofit_apps}}

\begin{figure*}
\centering
\setlength\tabcolsep{-2pt}
\begin{tabular}{@{}c@{}}
\subfloat{\includegraphics[width=1\textwidth]{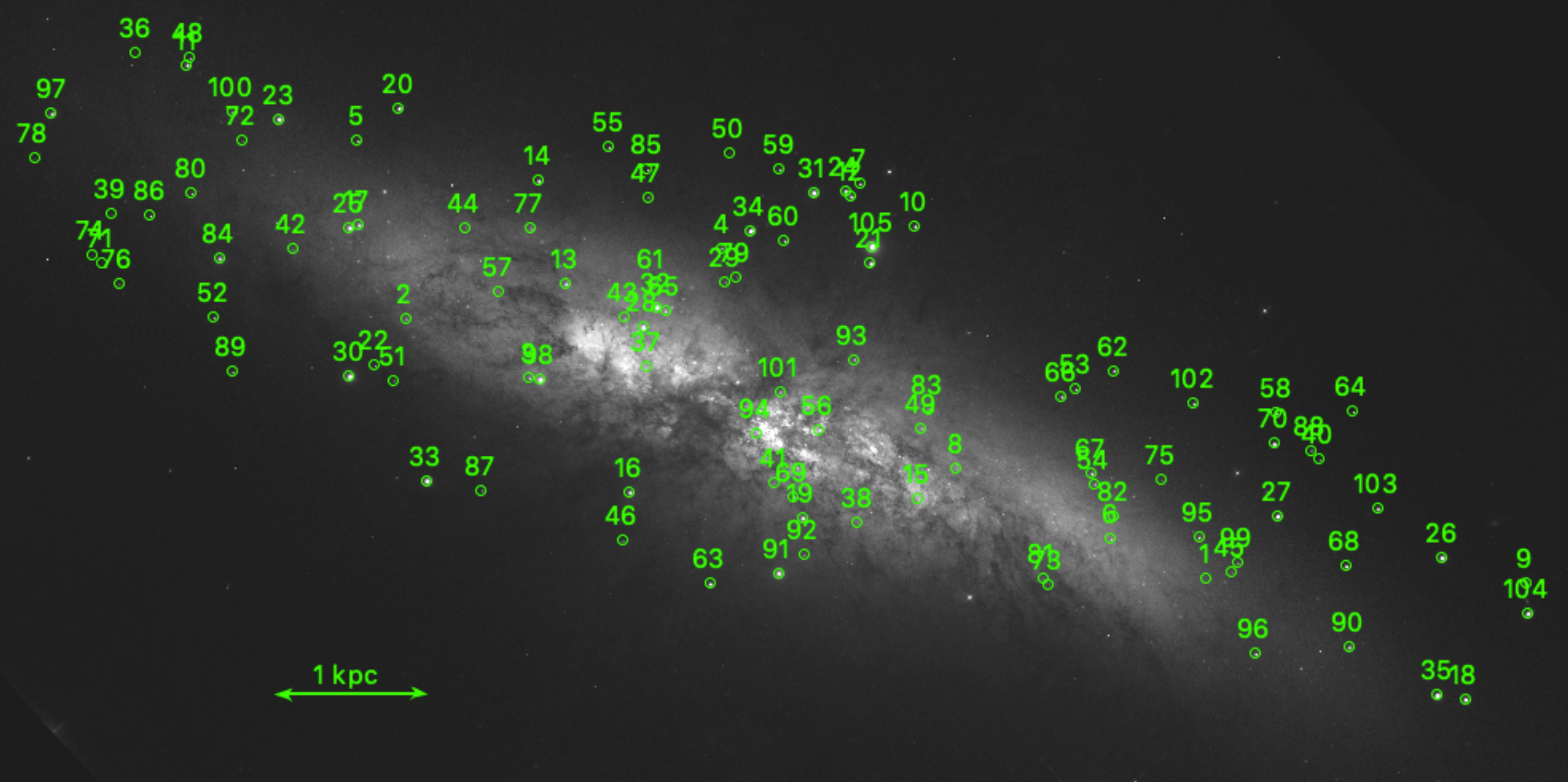}}
\end{tabular}\qquad 
\begin{tabular}{@{}c@{}}
\subfloat{\includegraphics[width=0.2\textwidth]{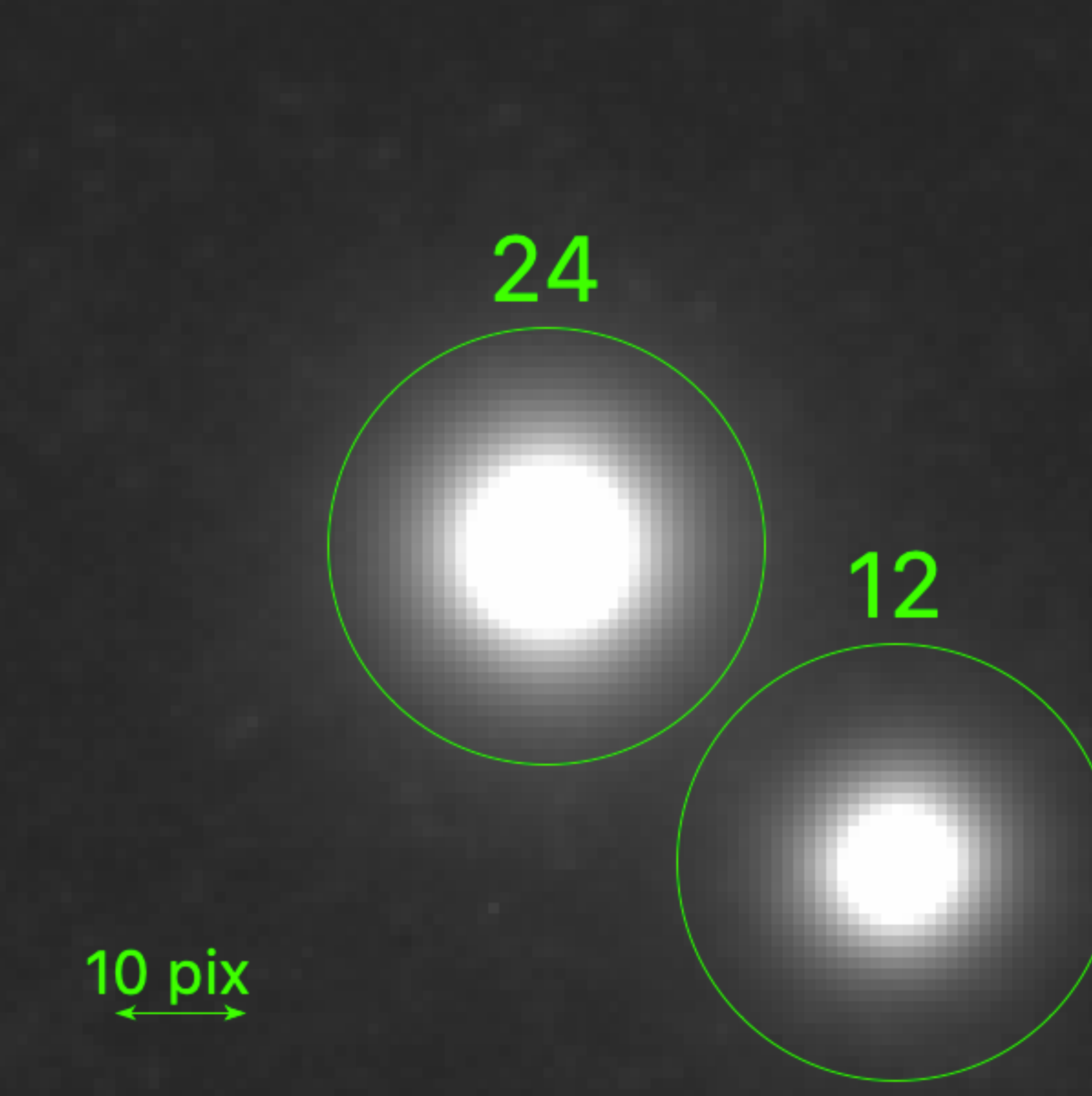}}
\subfloat{\includegraphics[width=0.2\textwidth]{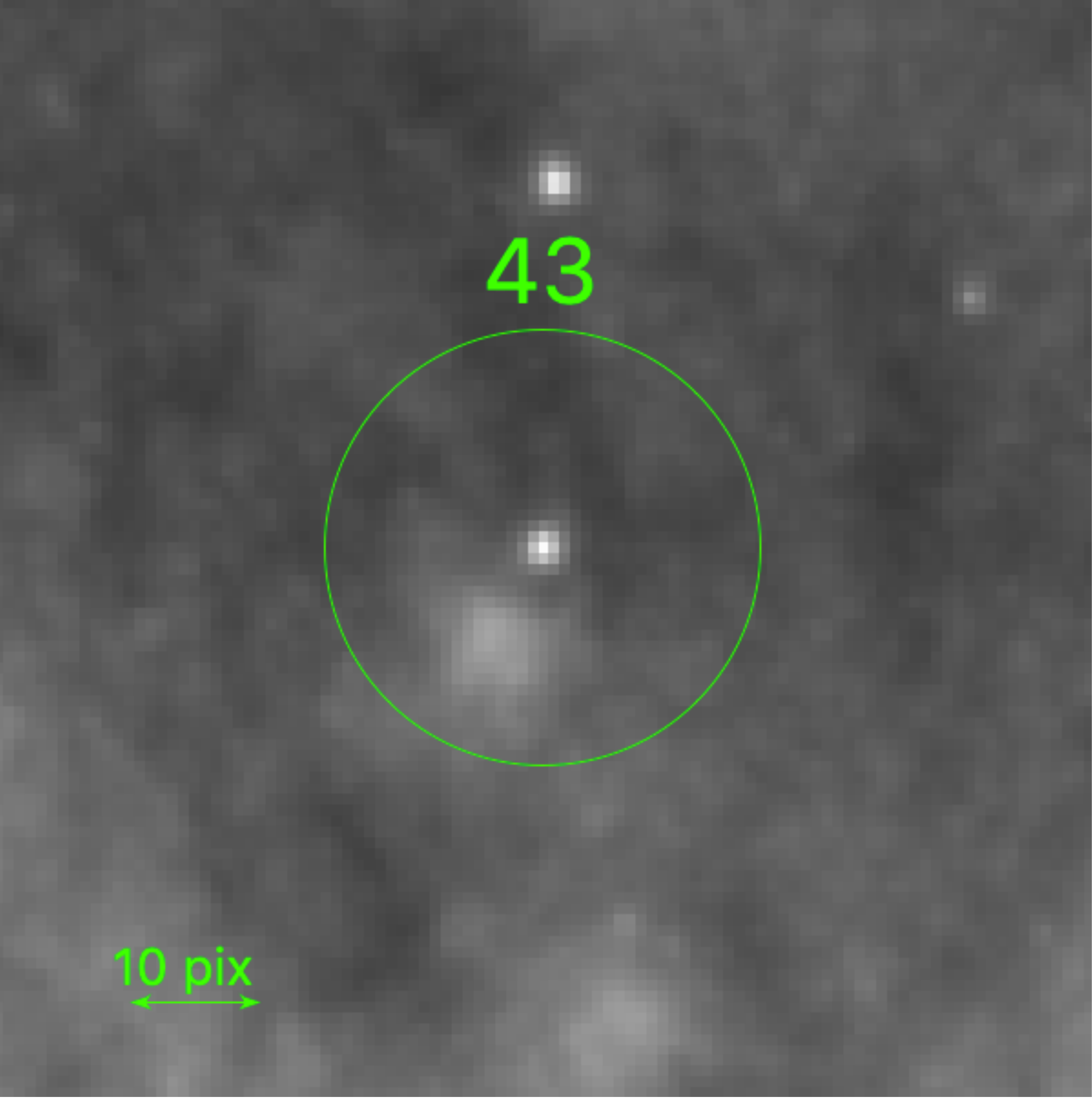}}
\subfloat{\includegraphics[width=0.2\textwidth]{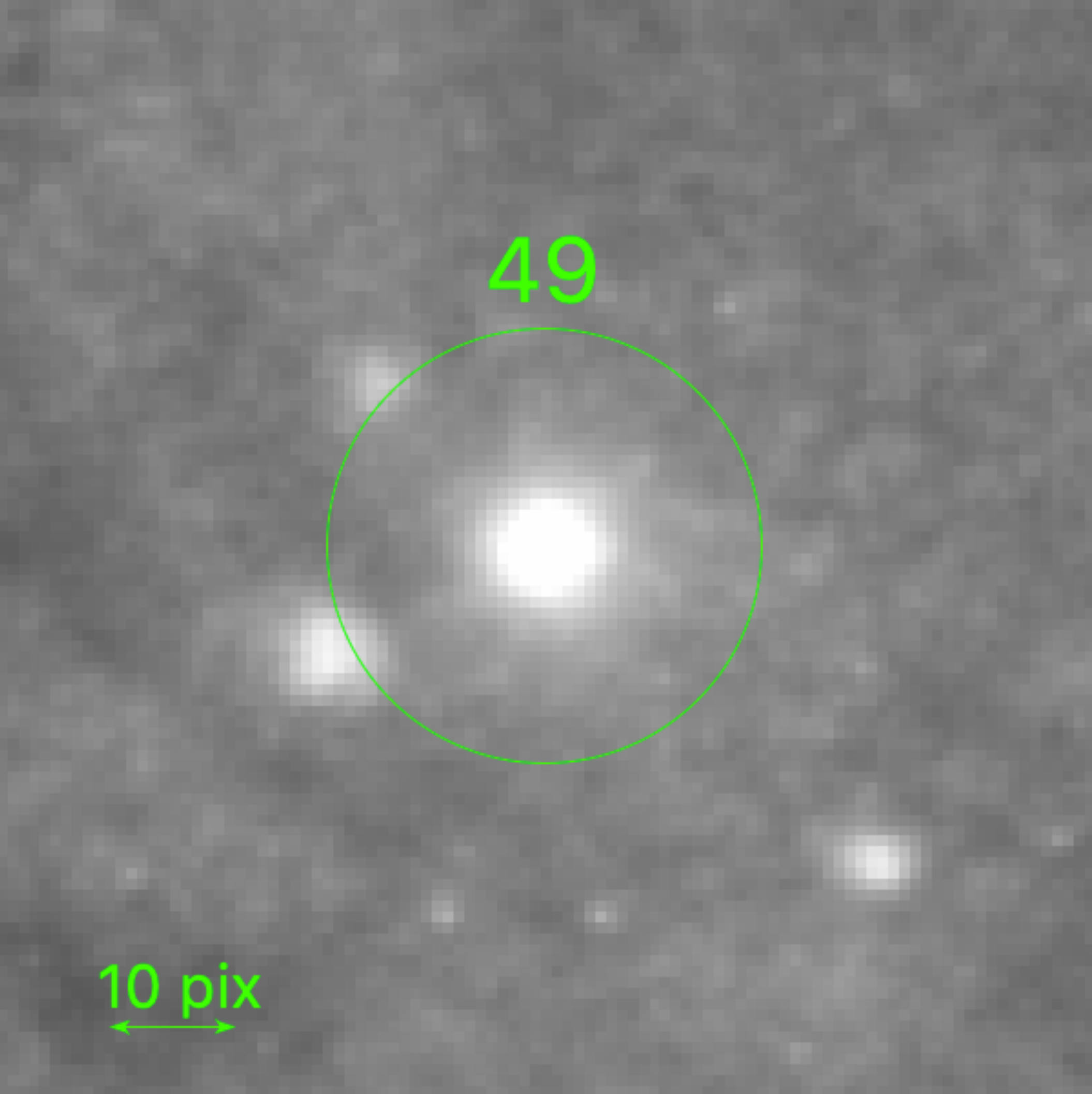}}
\subfloat{\includegraphics[width=0.2\textwidth]{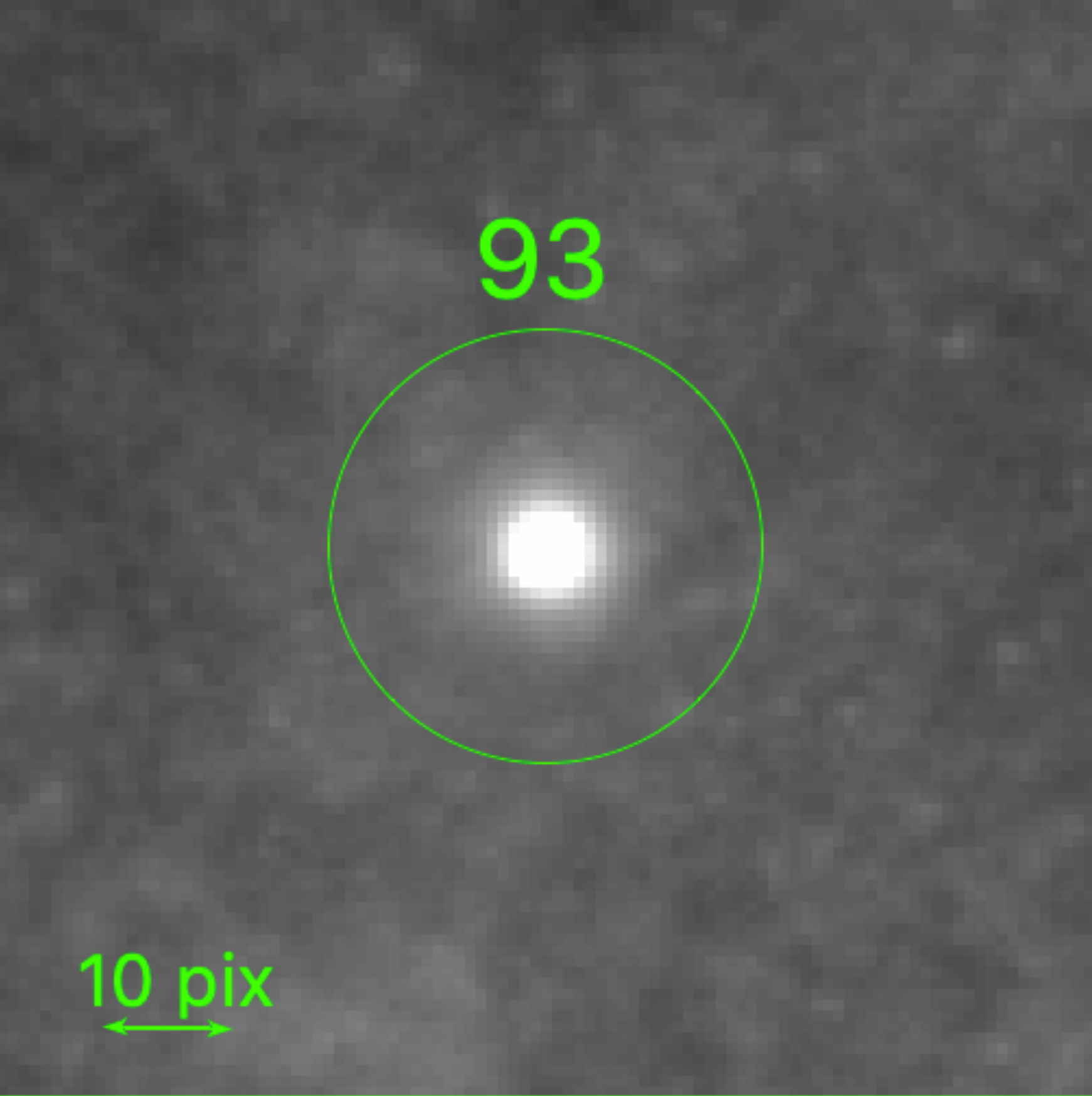}}
\subfloat{\includegraphics[width=0.2\textwidth]{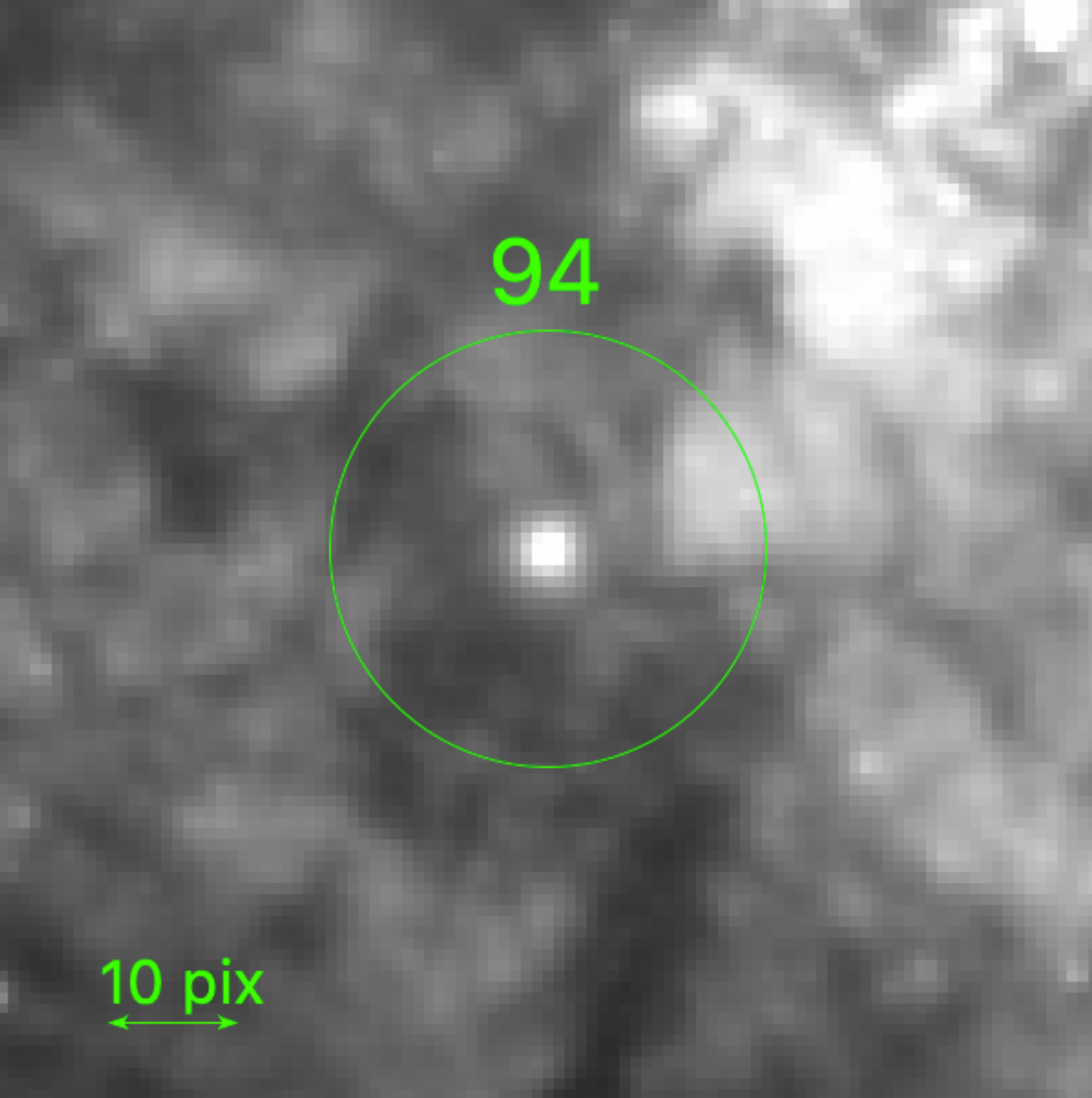}}
\end{tabular}
\caption{Top panel: 
Synthetic clusters (105 green circles) superposed on a real HST image, which corresponds to the F555W image. Bottom panels: zoom in of five clusters of the mock sample, in images of 101x101 pixels, centered in clusters S24, S43, S49, S93 and S94 (from left to right, respectively).  We show in the images bars indicating 1 kpc and 10 pixels, in the top and bottom panels, respectively. Throughout this work we use an image scale of 0.88~pc\,pixel$^{-1}$, which corresponds to the physical size of the HST/ACS pixels at the distance of M82~(3.63~Mpc).}
\label{fig:mock_sample}
\end{figure*}

In \citet{Cuevas2020,Cuevas2020b}, we have applied the techniques described in this algorithm, and compared our results with those obtained using the publicly available tools {\sc Galfit} and {\sc Ishape}. Our results for the sample of super star clusters in M82 were in agreement with the results obtained by {\sc Galfit} for Moffat-EFF models.  More specifically in \citet{Cuevas2020b}, we computed the derived parameters of the sample for Moffat-EFF models. In order to illustrate these techniques for King and Wilson models as well, we simulated a sample of 105 clusters following King, Wilson and, for the sake of completeness, we also simulated clusters following Moffat-EFF model profiles.  With the aim of testing the code in a realistic scenario, considering that typically, star clusters are embedded in crowded regions, we used an archive image as the background of our mock clusters. We used an image in F555W filter extracted from the HST Legacy Survey, provided by the Hubble Heritage Team, of the prototype starburst galaxy M82, which is a complex study case, due to its high crowding, high inclination angle and background gradient. In Fig \ref{fig:mock_sample}, we show a simulated image containing 105 clusters along with the zoom on sub-images around 5 of the clusters in the mock sample (the background of the mock sample has considerable gradient). The mock sample is based on the parameters set in Tab. \ref{tab:initialpars}, simulated several times for different central surface brightness.

\begin{deluxetable}{cccccc}[htbp]
\tablecaption{Set of initial parameters used to build the mock data with Moffat-EFF (M), King (K), and Wilson (W)}
\label{tab:initialpars}
\small\addtolength{\tabcolsep}{1 pt}
\tablehead{\colhead{$\gamma$ (M)} & \colhead{$r_d$ (M)} & \colhead{$W_0$ (K)} & \colhead{$r_0$ (K)} & \colhead{$W_0$ (W)} & \colhead{$r_0$} (W)\\
\colhead{} & \colhead{(pix)} & \colhead{} & \colhead{(pix)} & \colhead{} & \colhead{(pix)}\\
\colhead{(1)} & \colhead{(2)} & \colhead{(3)} & \colhead{(4)} & \colhead{(5)} & \colhead{(6)}}
\startdata
2.5  &  0.5  &  8  &  0.5  &    8.4  &  0.5 \\
2.7  &  2.7  &  8.2  &  0.9  &  8  &  0.9 \\
3.1  &  3.1  &  7.8  &  1.1  &  7.4  &  1.1 \\
3.1  &  3.5  &  6.4  &  1.9  &  6.2  &  1.9 \\
3.3  &  3.7  &  4.6  &  3.1  &  4.6  &  3.1 \\
3.5  &  4.3  &  3.8  &  4.7  &  5.2  &  4.9 \\
4.1  &  5.1  &  3.8  &  4.9  &  7.2  &  5.1 \\
\enddata
\tablecomments{Description of the columns: (1) Moffat-EFF shape parameter, (2) Moffat-EFF characteristic radius, (3) and (5), King and Wilson central dimensionless potential, (4) and (6), King Radius for King and Wilson models, respectively. Each pair of parameters was simulated for five different values of central surface brightness ($\rm mag/arcsec^{2}$): 18.5, 17.0, 16.3, 15.3, and 14. The varying galaxy background reaches values from 22.2 to 18.6 $\rm mag/arcsec^{2}$, with a median value of 21 $\rm mag/arcsec^{2}$, resulting in a heterogeneous sample with differences between the central surface brightness and the background value $\Delta \mu$ from 2 to 7.5 $\rm mag/arcsec^{2}$, with a median value of 4.7 $\rm mag/arcsec^{2}$.}
\end{deluxetable}

\subsection{Sample simulation}\label{Sec:mksample}

We have designed and implemented the subroutine {\sc mksample} to generate a mock sample, from user given coordinates and model type. 

Mock sample of clusters can be generated using our module {\sc mksample} from user given coordinates and model type.
{\sc mksample} generates 2D images following the projected profiles of the available models.  The synthetic profiles data are stored in a 2D matrix, which is generated following a similar procedure as that in the {\sc IRAF} task {\sc mkobjects}.  We draw particular attention to the geometrical features of the models. As in the case of {\sc mkobjects}, we consider the profiles of models with spherical symmetry. However, to reproduce the axysymmetry of some observed objects, we introduce an artificial ellipticity in the $x,y$ model coordinates as follows 
\begin{equation}
\begin{array}{l}
x_{mod}=x_t \cos((90-{\rm PA}) \frac{\pi}{180})+y_t\sin((90-{\rm PA})\frac{\pi}{180})\\
\\
y_{mod}=(-x_t \sin((90-{\rm PA}) \frac{\pi}{180})+y_t\cos((90-{\rm PA})\frac{\pi}{180}))/{\rm AR}\\
\\
{\rm where}\\
\quad \quad \quad \quad \quad \quad \quad x_t=x_i+1-x_c\\
\\
\quad \quad \quad \quad \quad \quad \quad  y_t=y_j+1-y_c            
\end{array}
\end{equation}
with PA and AR (AR=b/a, with a and b, the semi-major and semi-minor axis, respectively), the position angle and axis ratio, respectively.  $x_i$ and $y_j$ are the $(i,j)$ coordinates in the $x$ and $y$ direction of the image, and $x_c$, and $y_c$ the centers of each object. We draw attention to the mock elliptic clusters, whose parameters are well recovered using spherical models, since as we have shown in Appendix A in \citet{Cuevas2020}, for the most elongated cluster in M82, M82-F, the ellipticity does not considerably affects the integrated profile obtained from the isophotal fitting.

\begin{figure*}[htbp]
\centering
\setlength\tabcolsep{-5pt}
\setlength\extrarowheight{-10pt}
\begin{tabular}{@{}c@{}}
\subfloat{\includegraphics[width=0.34\textwidth]{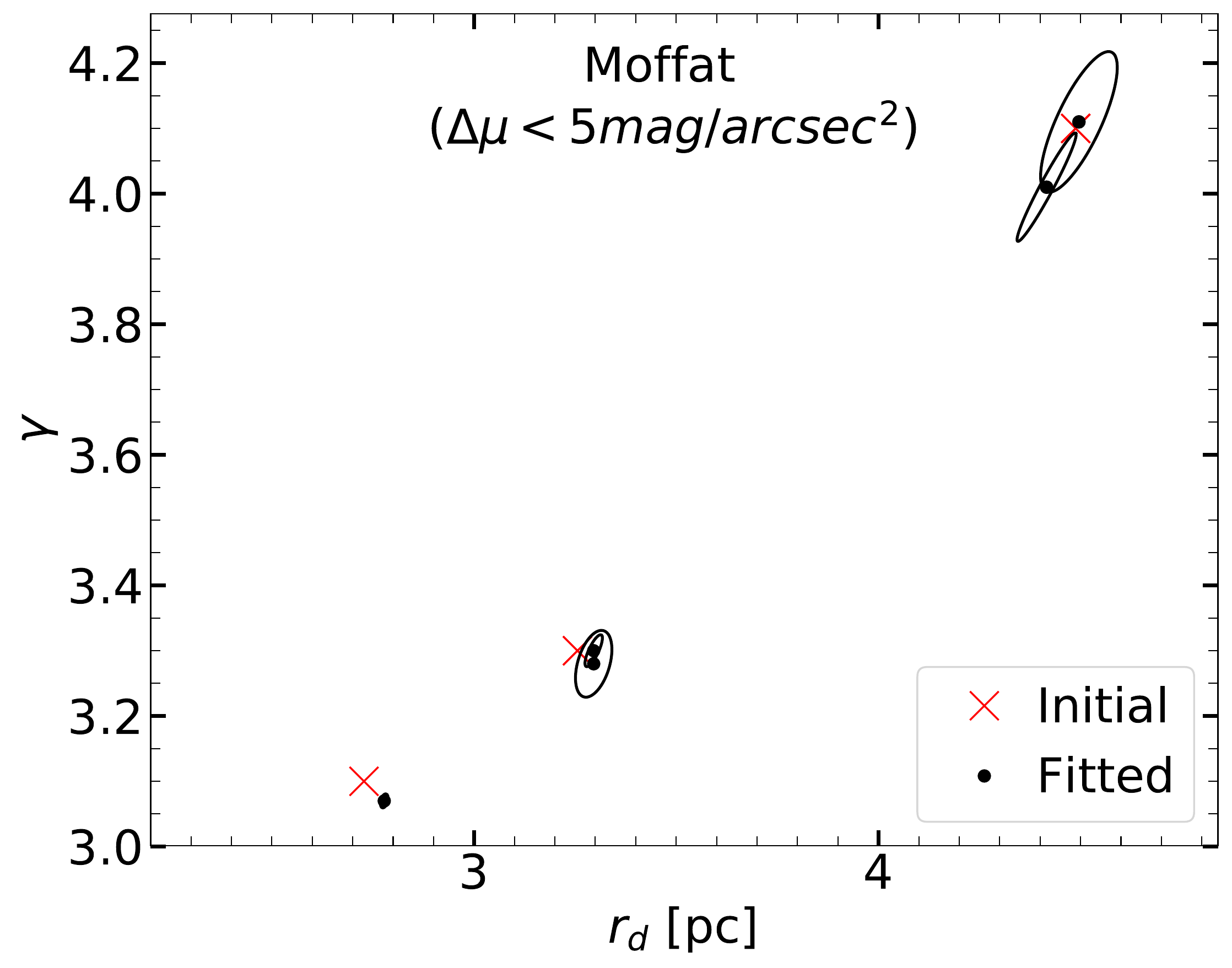}}
\subfloat{\includegraphics[width=0.34\textwidth]{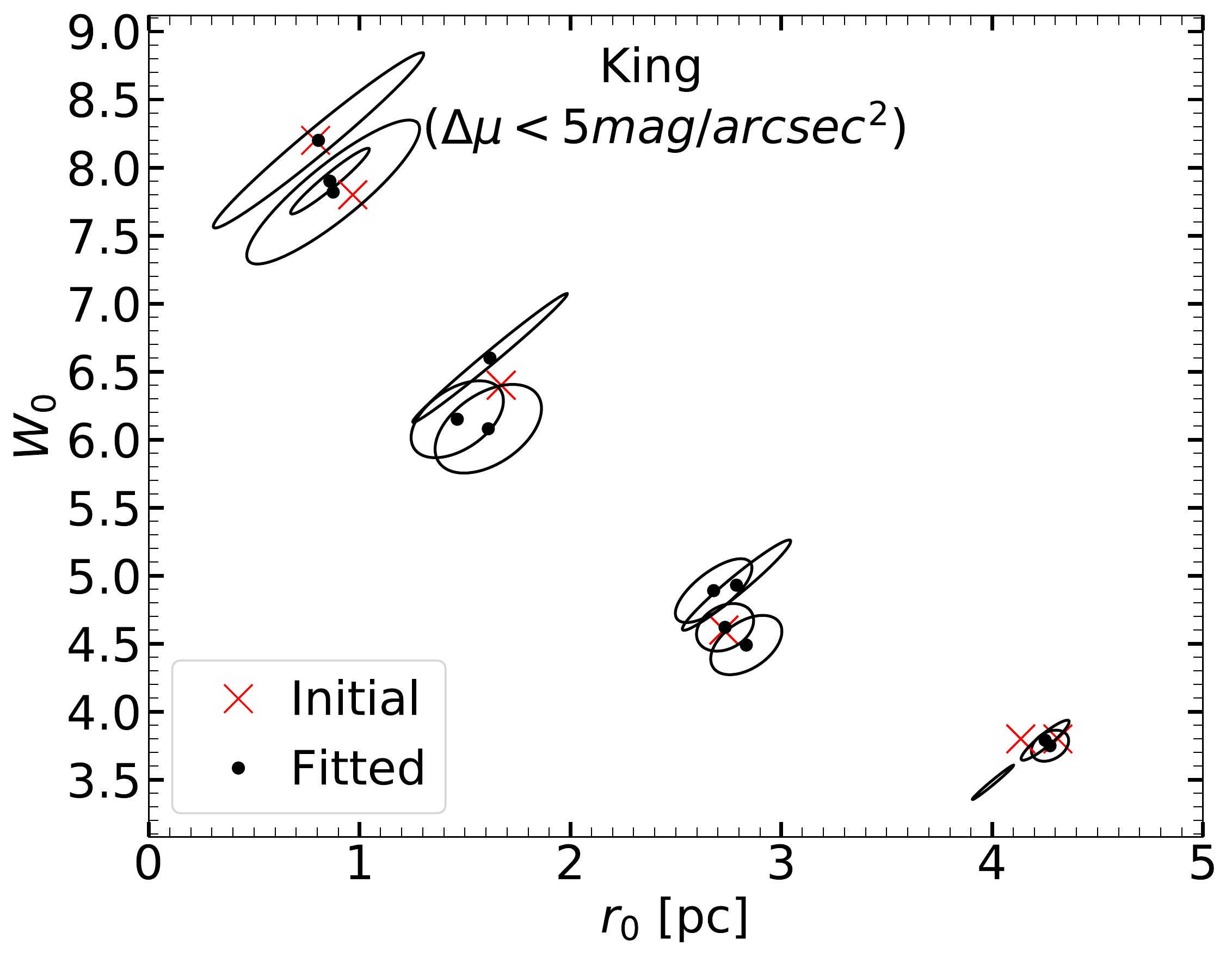}}
\subfloat{\includegraphics[width=0.34\textwidth]{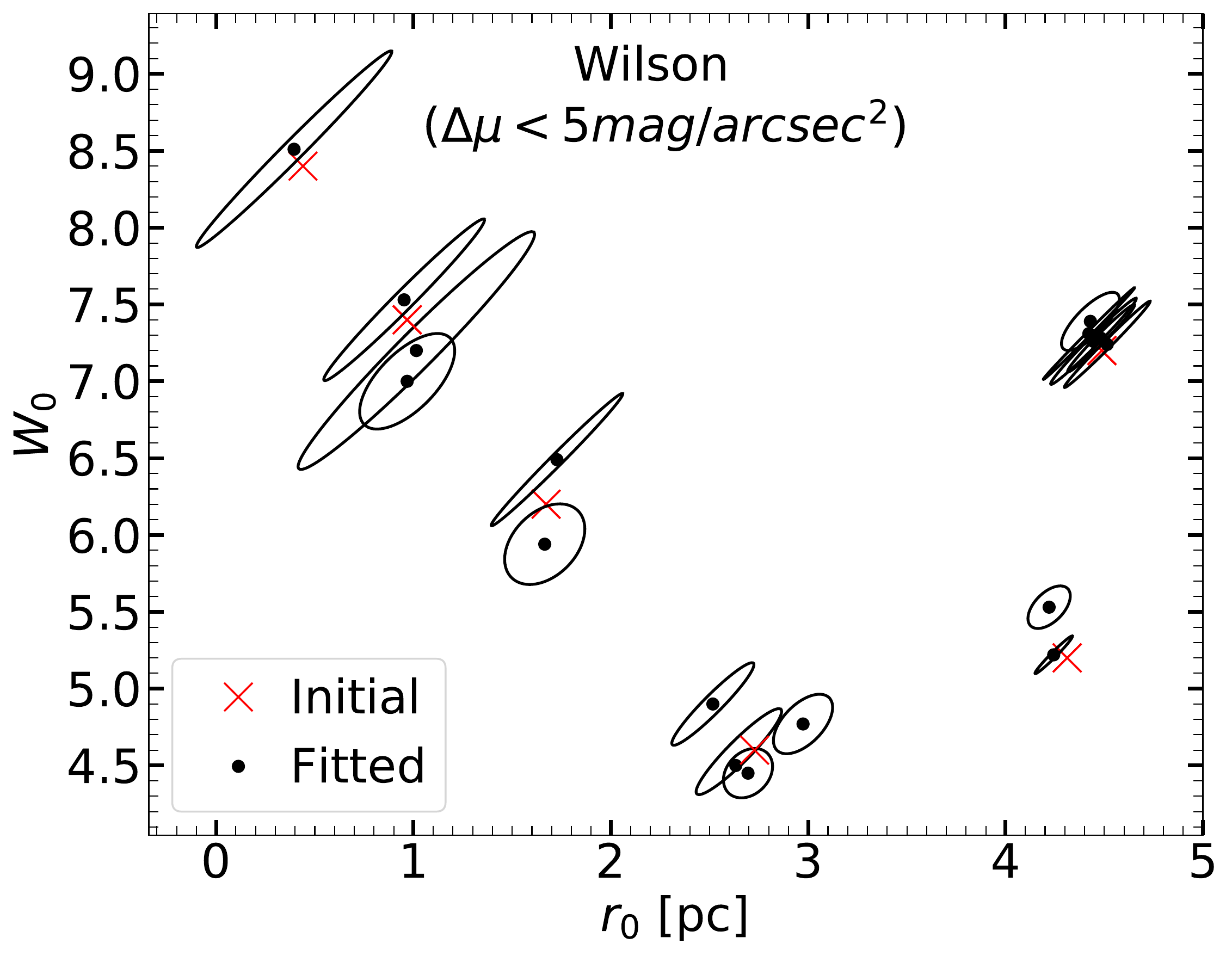}}\\
\subfloat{\includegraphics[width=0.34\textwidth]{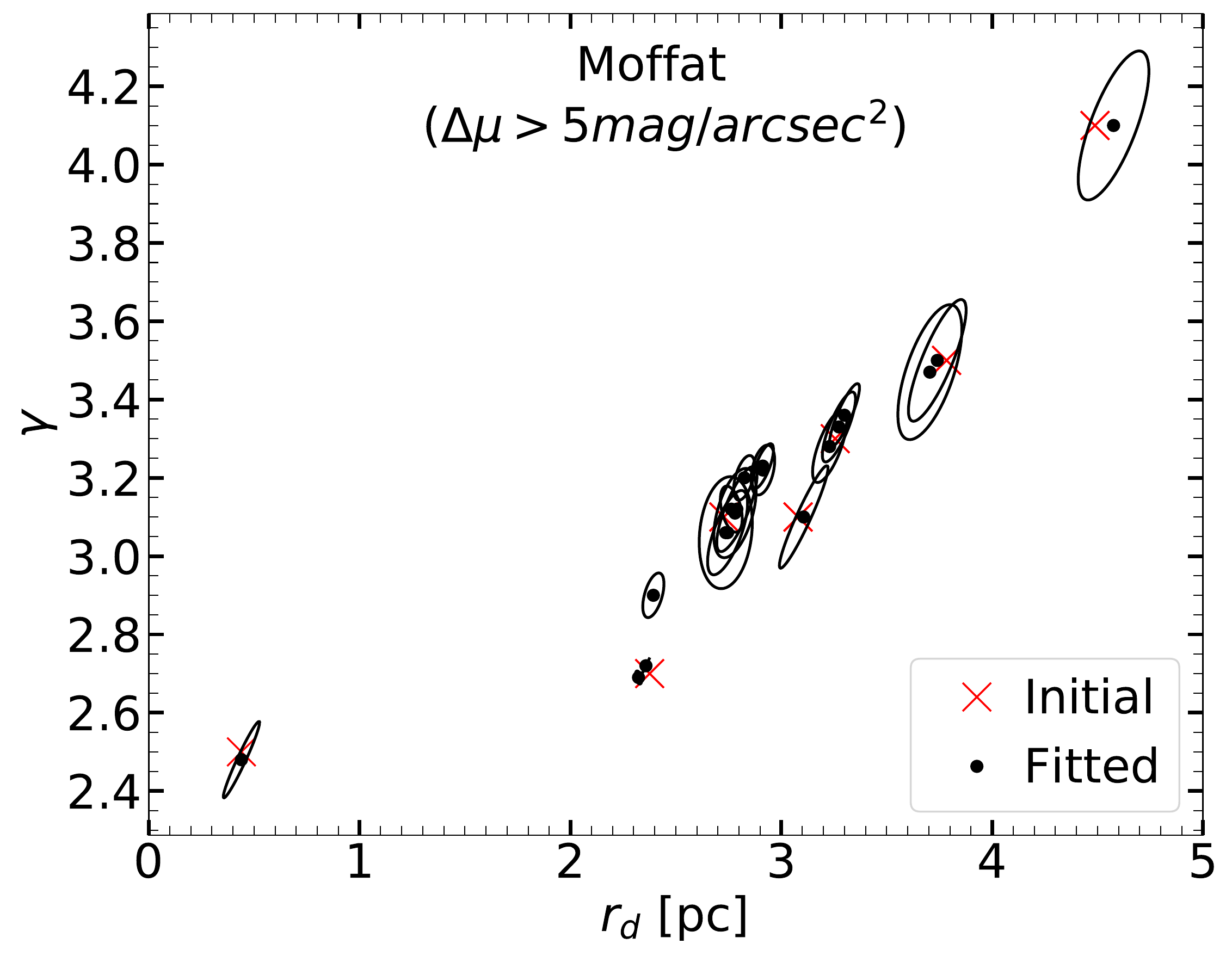}}
\subfloat{\includegraphics[width=0.34\textwidth]{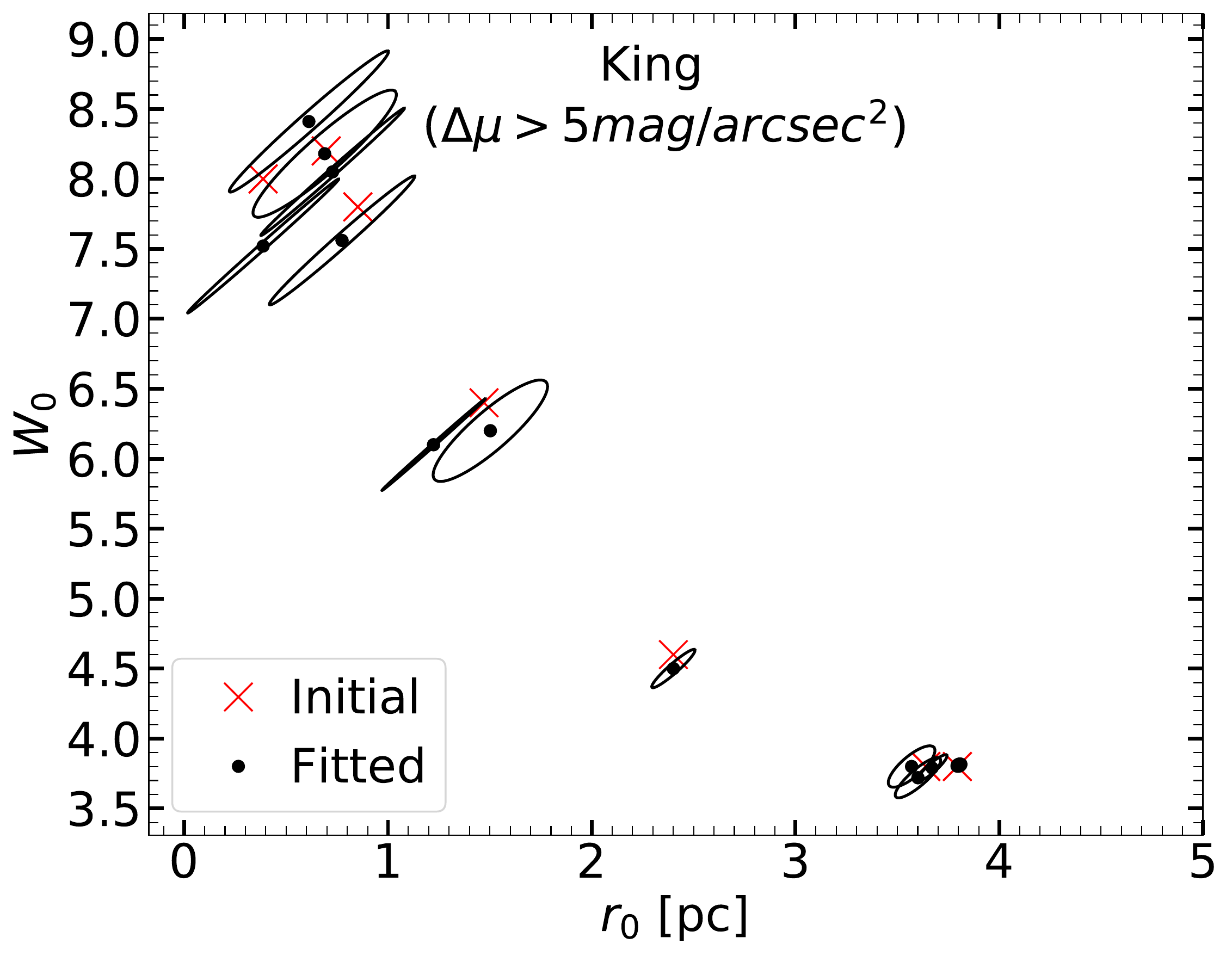}}
\subfloat{\includegraphics[width=0.34\textwidth]{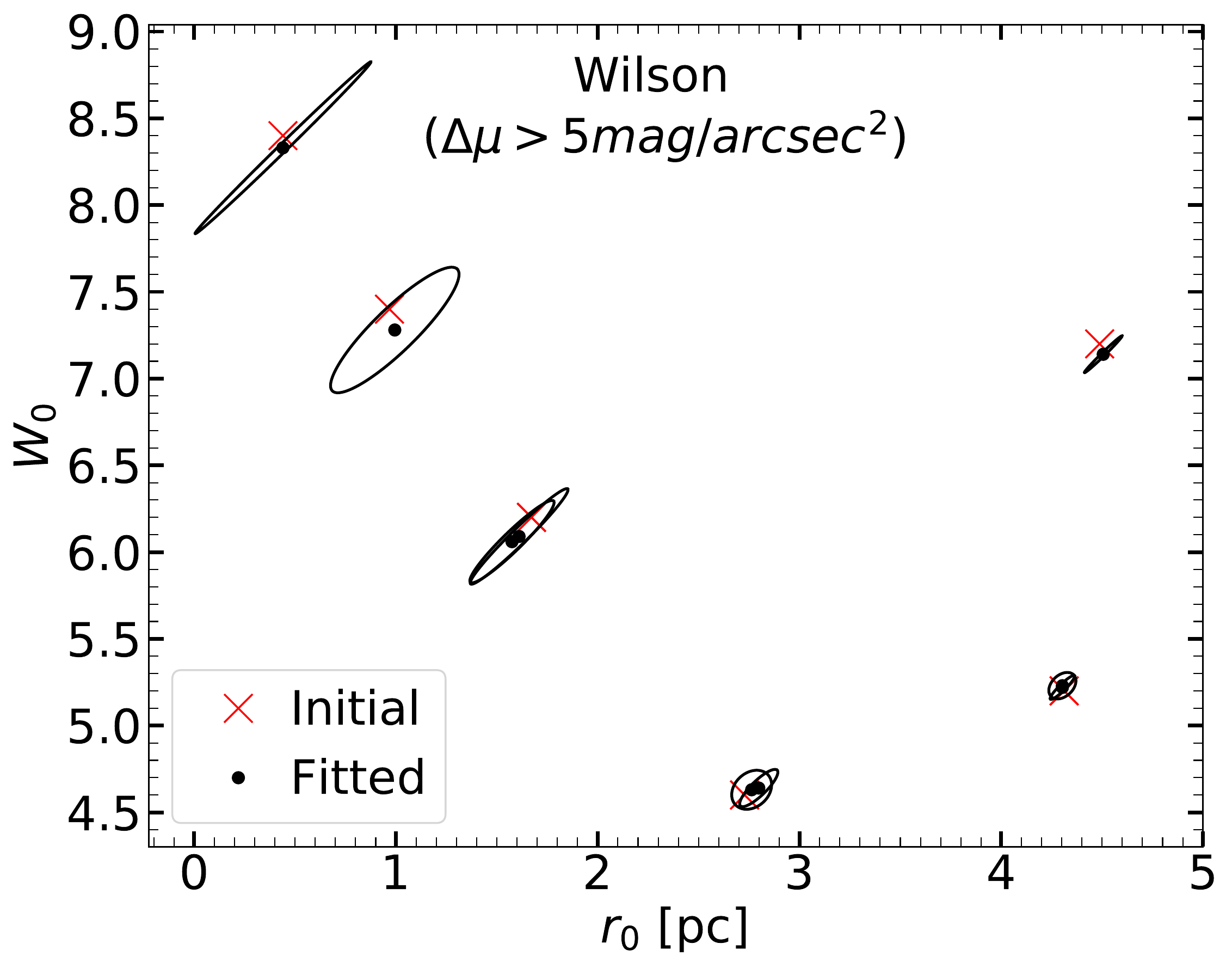}}
\end{tabular}
\caption{Initial conditions of the mock simulation, along with the values obtained by nProFit, with the corresponding 1$-\sigma$ confidence intervals for Moffat-EFF (left-most panels), King (middle panels), Wilson (right-most panels) models for faint (top panels) and bright (bottom panels) clusters.}
\label{fig:mock_ini}
\end{figure*}

\begin{figure*}[htbp]
\includegraphics[width=\textwidth]{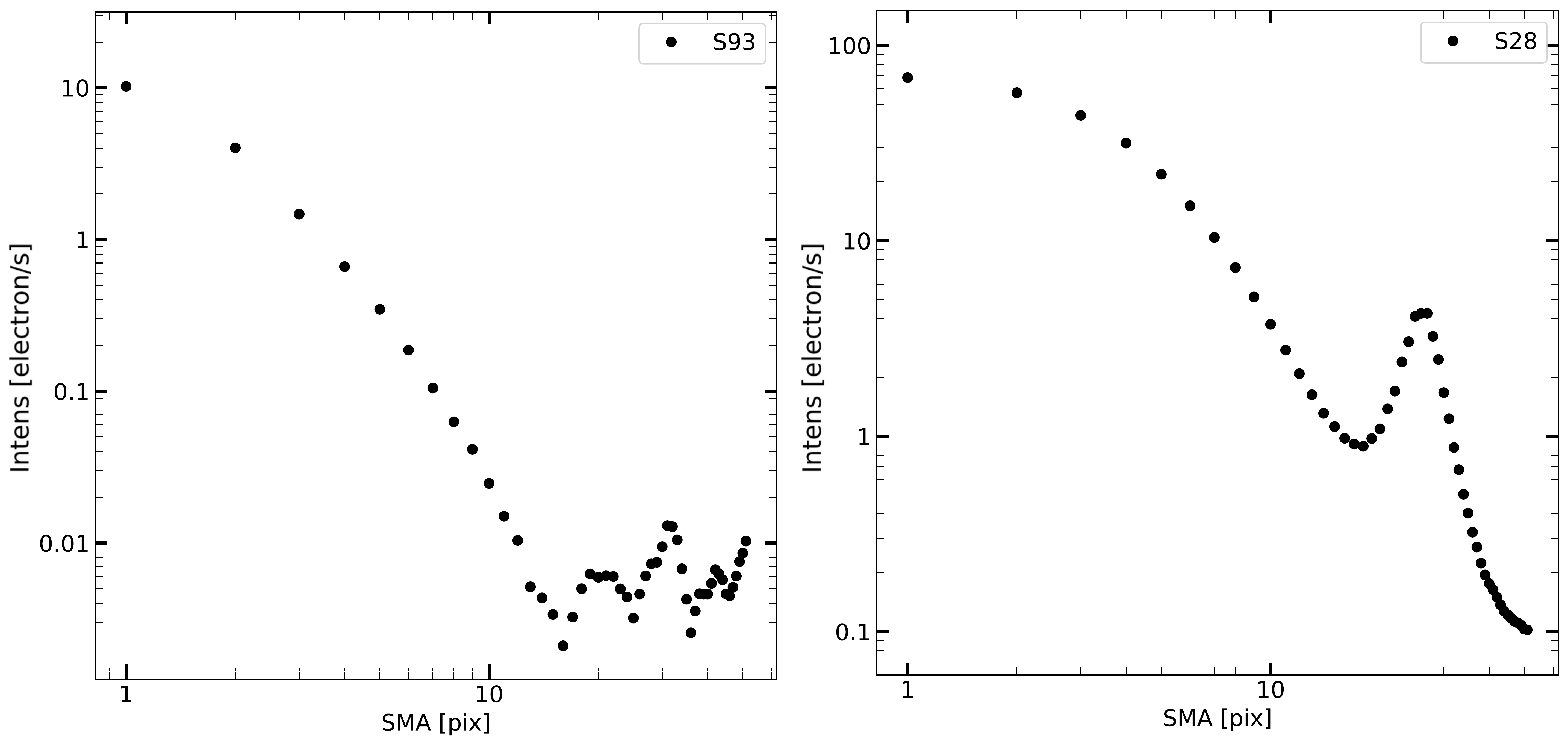}
\caption{Azimuthally averaged surface brightness profiles computed by {\sc nProFit} through the ellipse IRAF task for the objects S93 and S28, in the left and right panels, respectively.}
\label{fig:sbps}
\end{figure*}

\begin{figure*}
\includegraphics[width=\textwidth]{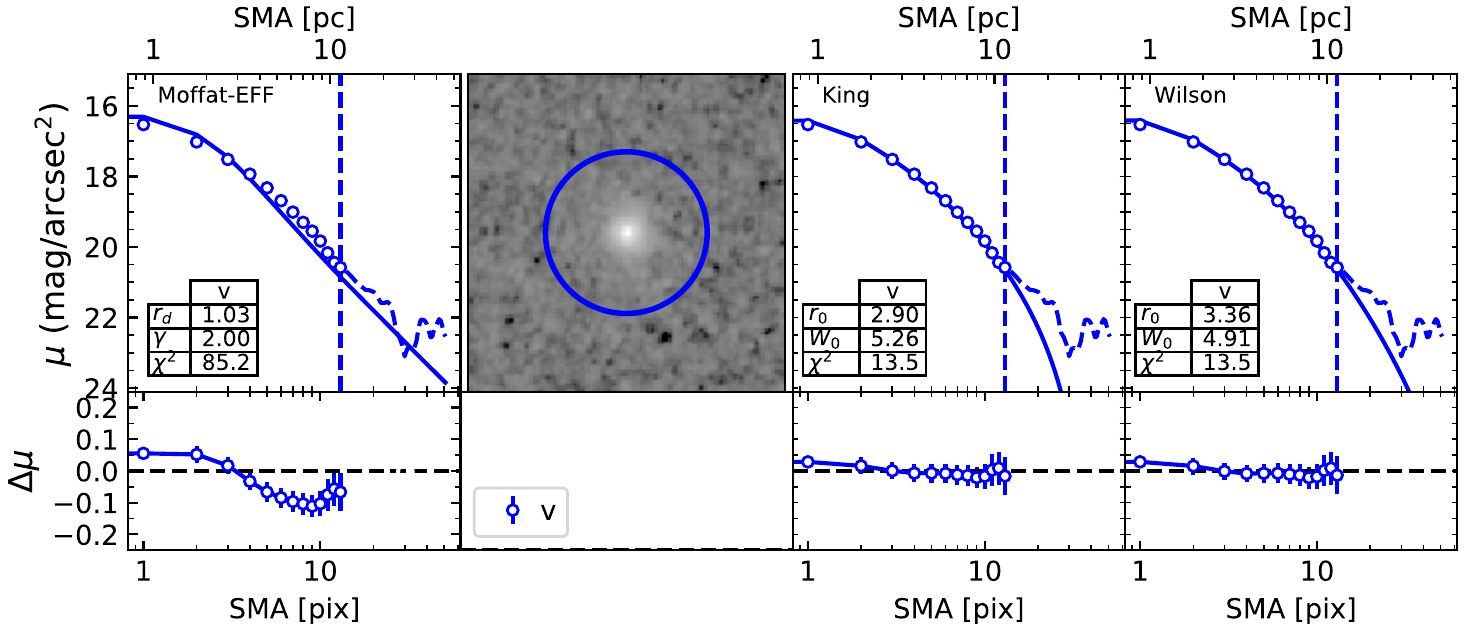}
\caption{Dynamical models (Moffat-EFF, King and Wilson) fitting performed by {\sc nProFit} to the surface brightness profile of a cluster in the mock sample generated with mksample (top panel). Fitting residuals (bottom panel). The fitting radius is shown with the vertical dashed line, the observed SBP with the empty dots and the fitted models with solid lines.}
\label{fig:prof_fit}
\end{figure*}

\begin{deluxetable*}{cccccccccccccrrc}
\tablecaption{Set of initial and fitted parameters obtained by nProFit for the mock sample}
\tablewidth{700pt}
\tabletypesize{\footnotesize}
\addtolength{\tabcolsep}{-3pt}
\tablehead{\colhead{ID} & \colhead{$\rm I_M$} & \colhead{$\rm O_M$}  & \colhead{$\rm \Delta \mu$} & \colhead{$\rm r_{0,i}$} & \colhead{$\rm r_{0,f}$} & \colhead{$\gamma$,$\rm W_0$} & \colhead{$\gamma$,$\rm W_0$} (f) & \colhead{$\rm R_{h,i}$} & \colhead{$\rm R_{h,f}$} & \colhead{$\rm R_c$} & \colhead{$\rm \log M$} & \colhead{$\rm \log \rho_0$} & \colhead{$\rm r_{t}$} & \colhead{$\rm \sigma_0$} & \colhead{$\rm \log E_b$}\\
\colhead{} & \colhead{} & \colhead{} & \colhead{$\rm \frac{mag}{arcsec^2}$} & \colhead{(pc)} & \colhead{(pc)}  & \colhead{} & \colhead{} & \colhead{(pc)} & \colhead{(pc)}  & \colhead{(pc)}  & \colhead{($\rm M_\odot$)}  &  \colhead{($\frac{\rm M_\odot}{pc^3}$)} & \colhead{(pc)} & \colhead{(km/s)} & \colhead{(ergs)}}
\decimalcolnumbers
\startdata
S5	 &  M	 & M	 & 3.5$\pm0.39$	 & 	3.3	 & 	3.1$_{-0.32}^{+0.37}$	 & 	3.3	 & 	3.0$_{-0.29}^{+0.31}$	 & 	4.5	 & 	4.8$_{-0.65}^{+0.75}$	 & 	2.1$_{-0.10}^{+0.13}$	 & 	5.5$_{-0.08}^{+0.07}$	 & 	3.0$_{-0.09}^{+0.10}$	 & 	11.4$_{-0.43}^{+0.48}$	 & 	3.1$_{-0.66}^{+0.69}$	 & 	46.0$_{-0.61}^{+0.68}$\\
S7	 &  M	 & M	 & 4.7$\pm0.38$	 & 	4.5	 & 	4.3$_{-0.39}^{+0.45}$	 & 	4.1	 & 	3.7$_{-0.35}^{+0.41}$	 & 	4.3	 & 	4.3$_{-0.81}^{+0.94}$	 & 	2.6$_{-0.10}^{+0.11}$	 & 	5.6$_{-0.02}^{+0.03}$	 & 	3.0$_{-0.08}^{+0.09}$	 & 	5.4$_{-0.52}^{+0.61}$	 & 	3.6$_{-0.72}^{+0.78}$	 & 	43.5$_{-0.74}^{+0.86}$\\
S11	 &  M	 & M	 & 5.5$\pm0.23$	 & 	3.1	 & 	3.1$_{-0.20}^{+0.22}$	 & 	3.1	 & 	3.0$_{-0.16}^{+0.13}$	 & 	4.9	 & 	4.9$_{-0.41}^{+0.43}$	 & 	2.1$_{-0.07}^{+0.10}$	 & 	6.0$_{-0.03}^{+0.01}$	 & 	3.5$_{-0.06}^{+0.06}$	 & 	21.8$_{-0.26}^{+0.26}$	 & 	5.3$_{-0.51}^{+0.51}$	 & 	46.6$_{-0.36}^{+0.36}$\\
S18	 &  M	 & M	 & 6.5$\pm0.16$	 & 	3.1	 & 	3.2$_{-0.16}^{+0.18}$	 & 	3.1	 & 	3.0$_{-0.13}^{+0.08}$	 & 	4.9	 & 	4.8$_{-0.32}^{+0.33}$	 & 	2.1$_{-0.06}^{+0.10}$	 & 	6.3$_{-0.02}^{+0.04}$	 & 	3.8$_{-0.04}^{+0.05}$	 & 	31.5$_{-0.21}^{+0.20}$	 & 	7.6$_{-0.45}^{+0.44}$	 & 	47.0$_{-0.29}^{+0.28}$\\
S31	 &  M	 & M	 & 7.3$\pm0.10$	 & 	2.7	 & 	2.8$_{-0.06}^{+0.06}$	 & 	3.1	 & 	3.0$_{-0.05}^{+0.04}$	 & 	4.3	 & 	4.3$_{-0.12}^{+0.12}$	 & 	1.9$_{-0.02}^{+0.03}$	 & 	6.6$_{-0.00}^{+0.01}$	 & 	4.3$_{-0.02}^{+0.02}$	 & 	18.6$_{-0.08}^{+0.07}$	 & 	11.6$_{-0.28}^{+0.27}$	 & 	47.7$_{-0.11}^{+0.10}$\\
S39	 &  K	 & K	 & 3.5$\pm0.67$	 & 	1.7	 & 	1.9$_{-3.05}^{+2.75}$	 & 	6.4	 & 	6.2$_{-0.79}^{+1.71}$	 & 	3.6	 & 	3.2$_{-4.80}^{+0.94}$	 & 	2.9$_{-0.70}^{+1.38}$	 & 	5.9$_{-0.11}^{+0.45}$	 & 	2.7$_{-0.17}^{+0.40}$	 & 	38.5$_{-2.24}^{+2.55}$	 & 	2.0$_{-0.22}^{+0.27}$	 & 	46.7$_{-0.06}^{+0.08}$\\
S49	 &  K	 & K	 & 2.5$\pm0.37$	 & 	4.3	 & 	4.0$_{-1.05}^{+1.05}$	 & 	3.8	 & 	5.3$_{-0.72}^{+0.98}$	 & 	4.5	 & 	5.1$_{-1.96}^{+0.33}$	 & 	2.9$_{-0.62}^{+0.81}$	 & 	6.0$_{-0.40}^{+0.49}$	 & 	2.9$_{-0.30}^{+0.43}$	 & 	49.2$_{-0.74}^{+0.80}$	 & 	5.3$_{-0.05}^{+0.06}$	 & 	46.7$_{-0.13}^{+0.13}$\\
S55	 &  K	 & K	 & 5.3$\pm0.26$	 & 	4.1	 & 	5.0$_{-0.36}^{+0.70}$	 & 	3.8	 & 	3.4$_{-0.81}^{+0.38}$	 & 	4.3	 & 	4.3$_{-0.86}^{+0.27}$	 & 	2.7$_{-0.63}^{+0.42}$	 & 	6.5$_{-0.45}^{+0.53}$	 & 	3.1$_{-0.44}^{+0.05}$	 & 	26.7$_{-0.30}^{+0.34}$	 & 	7.7$_{-0.04}^{+0.02}$	 & 	49.1$_{-0.01}^{+0.01}$\\
S58	 &  K	 & K	 & 6.1$\pm0.22$	 & 	0.8	 & 	0.8$_{-0.51}^{+0.59}$	 & 	8.2	 & 	8.4$_{-0.11}^{+0.14}$	 & 	5.2	 & 	5.3$_{-2.45}^{+0.14}$	 & 	3.0$_{-0.09}^{+0.11}$	 & 	7.7$_{-0.00}^{+0.03}$	 & 	4.0$_{-0.01}^{+0.01}$	 & 	67.2$_{-0.46}^{+0.55}$	 & 	3.9$_{-0.03}^{+0.04}$	 & 	49.4$_{-0.01}^{+0.01}$\\
S70	 &  K	 & K	 & 6.8$\pm0.15$	 & 	4.3	 & 	4.7$_{-0.35}^{+0.29}$	 & 	3.8	 & 	3.8$_{-0.37}^{+0.44}$	 & 	4.5	 & 	4.4$_{-0.52}^{+0.29}$	 & 	2.7$_{-0.32}^{+0.36}$	 & 	6.9$_{-0.27}^{+0.25}$	 & 	3.6$_{-0.16}^{+0.21}$	 & 	29.4$_{-0.22}^{+0.21}$	 & 	13.2$_{-0.02}^{+0.02}$	 & 	49.8$_{-0.01}^{+0.01}$\\
S76	 &  W	 & W	 & 3.8$\pm0.63$	 & 	4.3	 & 	4.7$_{-0.59}^{+2.01}$	 & 	5.2	 & 	5.4$_{-1.72}^{+0.57}$	 & 	3.5	 & 	3.5$_{-0.36}^{+0.36}$	 & 	2.8$_{-1.29}^{+0.65}$	 & 	5.4$_{-0.69}^{+1.02}$	 & 	2.4$_{-1.03}^{+0.10}$	 & 	165.6$_{-0.83}^{+1.72}$	 & 	3.2$_{-0.09}^{+0.09}$	 & 	45.1$_{-0.11}^{+0.33}$\\
S82	 &  W	 & W	 & 2.5$\pm0.39$	 & 	2.7	 & 	2.9$_{-1.45}^{+1.45}$	 & 	4.6	 & 	5.0$_{-0.78}^{+1.12}$	 & 	2.1	 & 	2.0$_{-0.26}^{+0.28}$	 & 	2.8$_{-0.76}^{+1.00}$	 & 	6.0$_{-0.21}^{+0.39}$	 & 	3.0$_{-0.21}^{+0.34}$	 & 	73.6$_{-1.22}^{+1.34}$	 & 	4.1$_{-0.08}^{+0.10}$	 & 	46.9$_{-0.05}^{+0.04}$\\
S88	 &  W	 & W	 & 5.6$\pm0.28$	 & 	1.7	 & 	1.8$_{-0.96}^{+0.54}$	 & 	6.2	 & 	6.1$_{-0.22}^{+0.41}$	 & 	1.7	 & 	1.5$_{-0.48}^{+0.47}$	 & 	2.9$_{-0.24}^{+0.34}$	 & 	6.7$_{-0.03}^{+0.11}$	 & 	3.5$_{-0.04}^{+0.09}$	 & 	117.9$_{-1.18}^{+0.82}$	 & 	4.6$_{-0.04}^{+0.05}$	 & 	47.6$_{-0.03}^{+0.02}$\\
S102	 &  W	 & W	 & 6.4$\pm0.16$	 & 	1.7	 & 	1.8$_{-0.39}^{+0.44}$	 & 	6.2	 & 	6.1$_{-0.20}^{+0.19}$	 & 	1.7	 & 	1.6$_{-0.28}^{+0.21}$	 & 	2.9$_{-0.18}^{+0.17}$	 & 	7.1$_{-0.02}^{+0.06}$	 & 	3.9$_{-0.04}^{+0.04}$	 & 	124.6$_{-0.54}^{+0.59}$	 & 	7.7$_{-0.03}^{+0.02}$	 & 	48.5$_{-0.01}^{+0.01}$\\
S103	 &  W	 & W	 & 6.9$\pm0.16$	 & 	2.7	 & 	2.9$_{-0.40}^{+0.42}$	 & 	4.6	 & 	4.7$_{-0.27}^{+0.28}$	 & 	2.1	 & 	2.0$_{-3.08}^{+3.63}$	 & 	2.7$_{-0.25}^{+0.26}$	 & 	6.7$_{-0.06}^{+0.10}$	 & 	3.8$_{-0.08}^{+0.08}$	 & 	64.5$_{-0.33}^{+0.35}$	 & 	9.6$_{-0.02}^{+0.02}$	 & 	48.6$_{-0.01}^{+0.01}$\\
\enddata
\tablecomments{Description of the columns:
(1) Simulated cluster ID. (2) Input model. (3) Output model. (4) Difference between central surface brightness and local background value.
(5) and (6) Initial and fitted scale radius $r_d$ for Moffat-EFF models and $r_0$ for King and Wilson models.  (7) and (8) Initial and fitted shape parameters $\gamma$ for Moffat-EFF models and $W_0$ for King and Wilson models.  (9) and (10) Initial and fitted half-light radii. (11) Core radius. (12) Logarithm of total profile mass. (13) Logarithm of central volume mass density. (14) Tidal radius. (15) Central velocity dispersion. (16) Logarithm of the binding energy.}
\end{deluxetable*}

Each profile is simulated as follows
\begin{equation}
{\rm sim(i,j)}=S_{0_{mod}}\,\, {\rm mod(r,par1,par2)+bg}
\end{equation}
\noindent with $r=\sqrt{x_{mod}^2+y_{mod}^2}$, mod the selected model (Moffat-EFF, Wilson or King) $S_{0_{mod}}$ the central surface brightness used for the simulation, par1, par2, the model parameters ($r_d$ and $\gamma$ for Moffat-EFF models, and $r_0$ and $W_0$ for King and Wilson models), bg is the simulated background value for each coordinate. For the sake of testing {\sc nProFit} in a realistic background scenario, we used a real archive image of a galaxy instead of the usual background image drawn from a Gaussian distribution with mean and sigma values provided by the user.  We used an image from the HST Legacy Survey made publicly available by the Hubble Heritage Team \citep{Mutchler2007} corresponding to the prototype starburst galaxy M82, located at a distance of 3.63 Mpc \citep{Freedman}. Such a galaxy represents an interest study case, ideal to determine the extents of the code, considering that it has a strong background gradient, a large number of clusters along its disk and nucleus (around 600) \citep{Mayyacat} and a high inclination degree 77$^\circ$ \citep{Mayya2005}. In addition to the realistic background conditions, the mock sample spans a wide range of central surface brightness values (spanning around 4.5~mag/arcsec$^2$, from 18.5~mag/arcsec$^2$ to 14~mag/arcsec$^2$, in 5 bins), initially set to test the accuracy of the code and its dependence on the surface brightness profiles of the clusters.

In Fig. \ref{fig:mock_sample}, we show the mock sample constituted by 105 clusters with the M82 galaxy image in the F555W as the background image.  We assumed a mass-to-light ratio of 1 (resembling that of a population of 2.2 Gyr in the F555W filter) and a zero-point magnitude of 25.779 mag. The initial structural parameters, were simulated, considering realistic values (following the values reported in \citet{Cuevas2020,Cuevas2020b}), resembling clusters in M82. To that aim, we performed our simulations, chosing the set of initial conditions in Fig. \ref{fig:mock_ini}, for Moffat-EFF, King, and Wilson models, with central surface brightness values between 14~mag/arcsec$^2$ and 18.5~mag/arcsec$^2$, and with background values between 18.6~mag/arcsec$^2$ and 22.2~mag/arcsec$^2$. We performed the fits using nProFit, and find that, the obtained parameters, are well within the $1-\sigma$ confidence intervals centred at the initial values, in the majority of cases, showing fitted values closer to the initial values for brighter clusters. Our mock sample is constituted by these 21 combinations of structural parameters (7 per each model, summarized in Tab. \ref{tab:initialpars}), along, with the 5 bins in surface brightness, resulting in a mock sample of 105 clusters, with each combination of structural parameters ($\gamma$ and $r_d$ for Moffat-EFF, and $W_0$ and $r_0$ for King and Wilson models), simulated for each one of the initial central surface brightness profiles. The mock sample clusters positions were drawn from a uniform distribution limited by the galaxy geometry.

\subsection{Structural parameters of the mock sample}

We run the {\sc nProFit} code over our synthetic data, and obtained in the first place the sub-images centered in each object.  Background subtracted images are subsequently obtained. 

In Fig. \ref{fig:sbps} we show examples of SBPs on our simulated image for two clusters. The intensities are given in units of raw counts per second (cps) on the background subtracted images. These are converted to mag\,arcsec$^{-2}$ units using the zeropoint and image scale given by the user. The profile semi major axis is given in units of pixels, which are converted to parsecs, using the image scale given by the user, to match the theoretical models units for fitting purposes.

\begin{figure}
    \centering
    \includegraphics[width=\columnwidth]{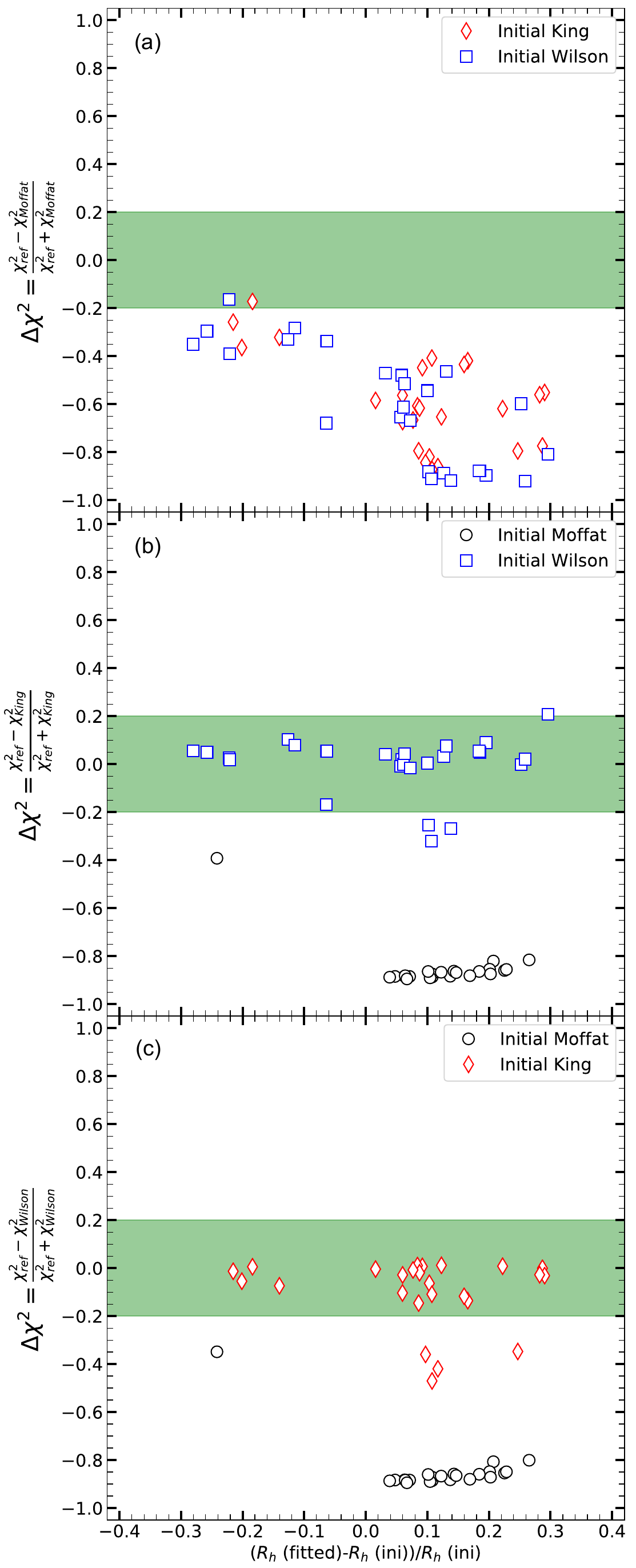}
    \caption{$\Delta \chi^2$ values to determine the best fitting model, setting as a model for comparison Moffat-EFF (a), King (b) and Wilson (c) versus percentage error for $\rm R_h$ fits performed by {\sc nProFit}, for the clusters initially fitted with the models shown in the figure legend. The green horizontal gaps represent a difference of 20\%, between the compared models, where the compared fits provide equally good results.}
    \label{fig:delta_chi}
\end{figure}

\begin{figure*}[ht]
\centering
\setlength\tabcolsep{-5pt}
\setlength\extrarowheight{-10pt}
\begin{tabular}{@{}c@{}}
\subfloat{\includegraphics[width=0.5\textwidth]{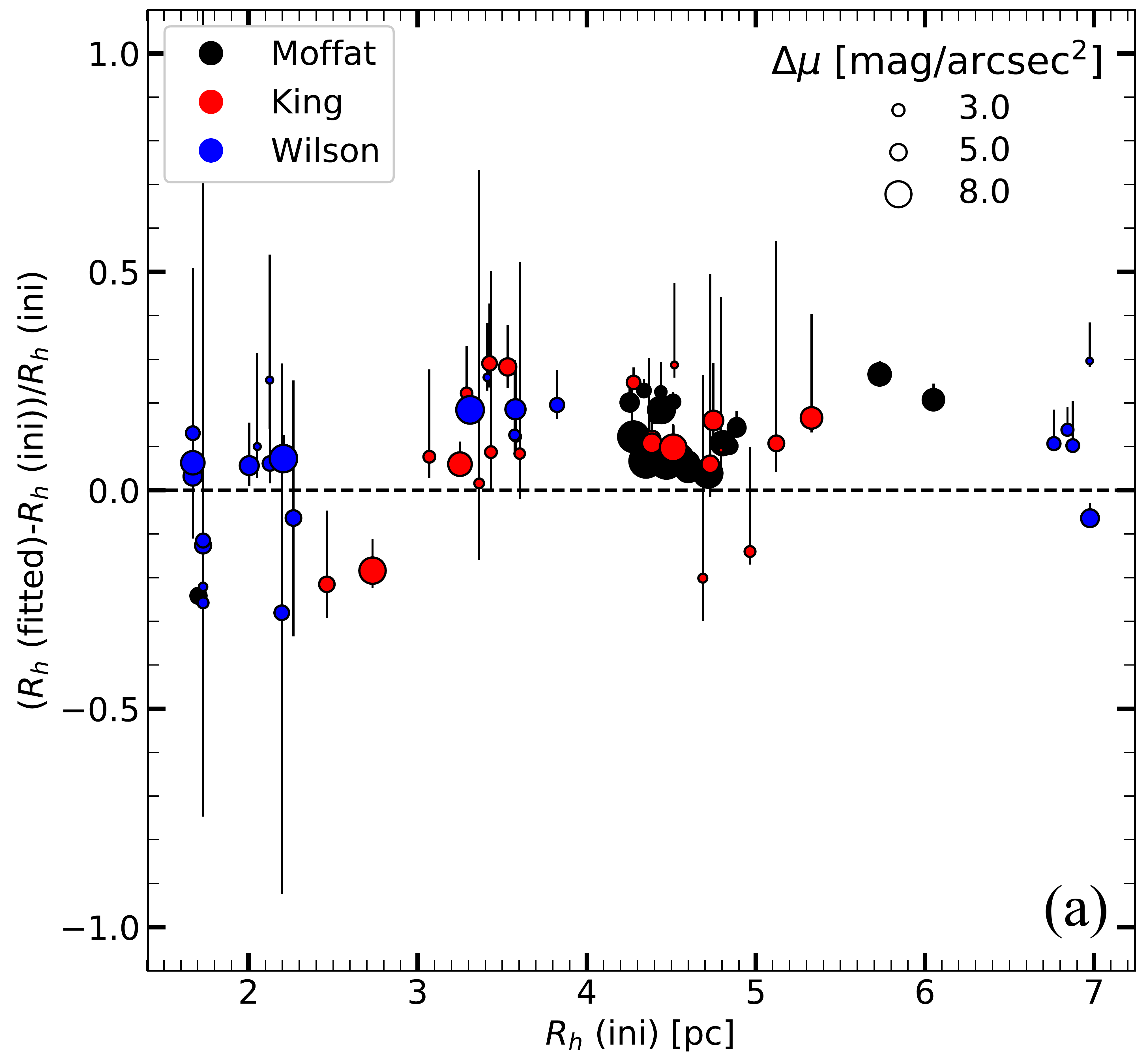}\label{fig:reff_nprofit}}
\subfloat{\includegraphics[width=0.5\textwidth]{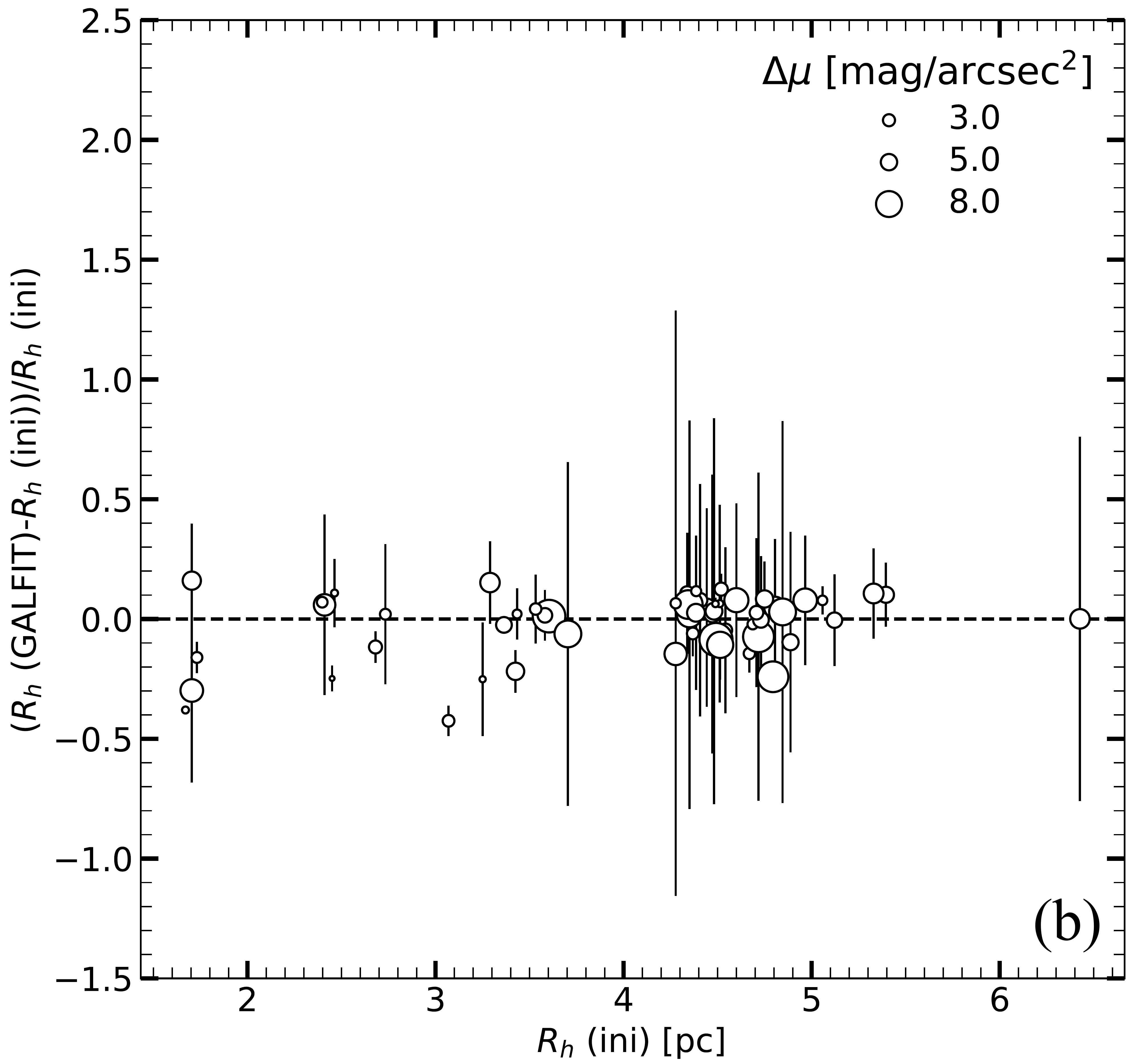}\label{fig:reff_galit}}\\
\subfloat{\includegraphics[width=0.5\textwidth]{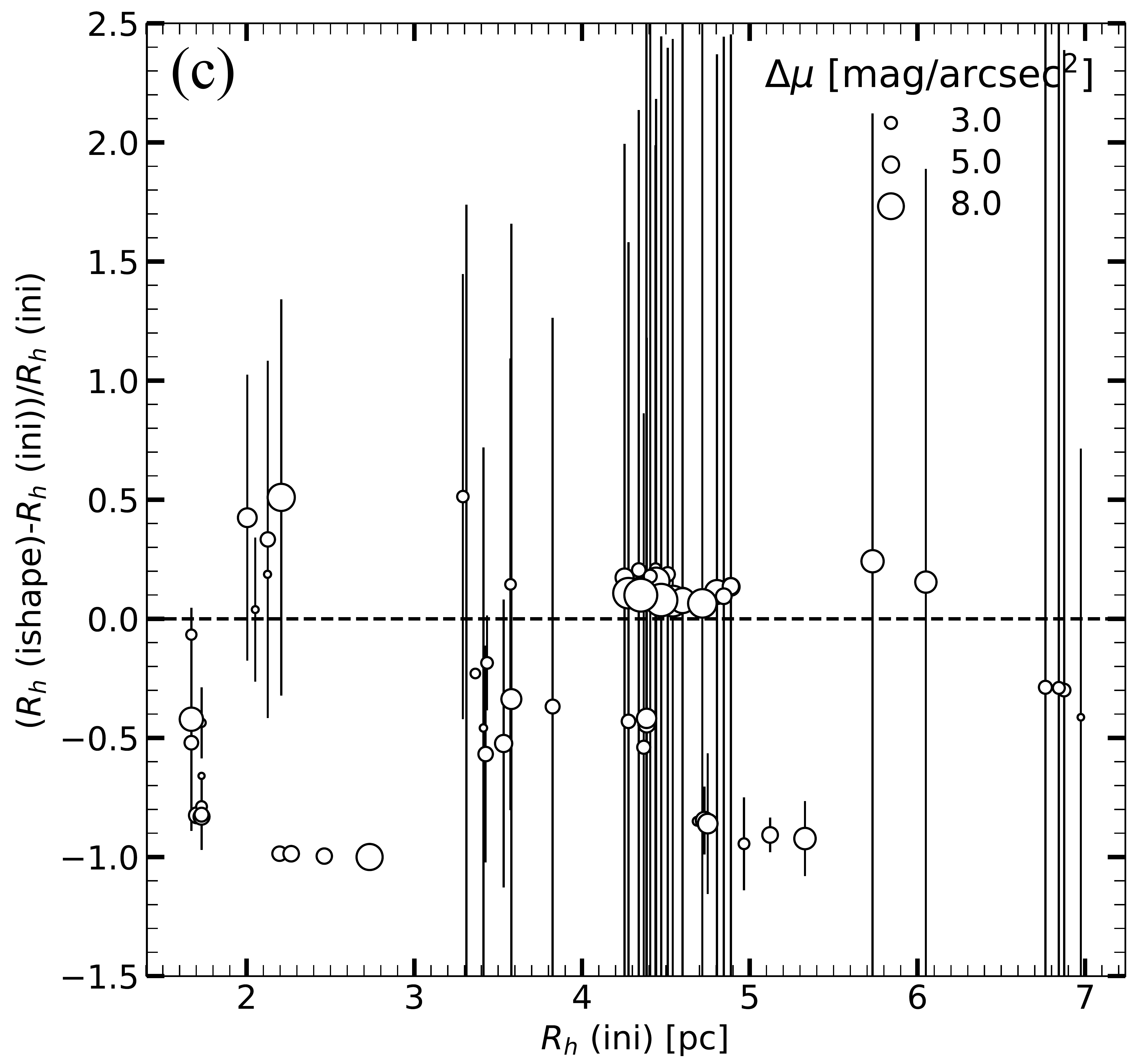}\label{fig:reff_ishape}}
\subfloat{\includegraphics[width=0.5\textwidth]{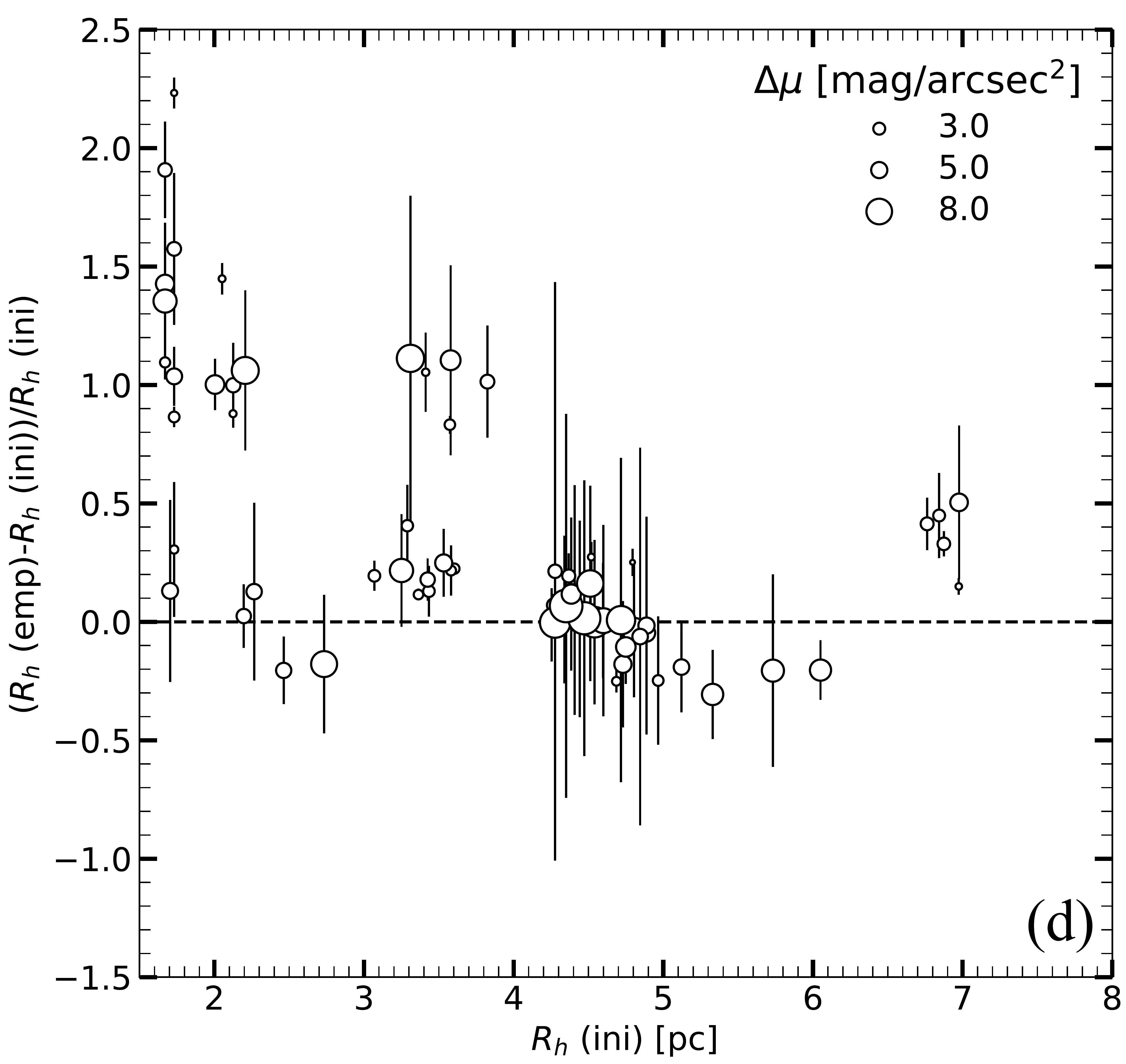}\label{fig:reff_emp}}
\end{tabular}
\caption{Percentage error for  $\rm R_h$ fits performed by {\sc nProFit} (a), {\sc Galfit} (b), {\sc Ishape}  (c), and from an empirical estimate (d)  versus the initial half-light radius, with the point sizes coded as a function of the difference between central surface brigthness and local background values for each cluster. The dotted horizontal line represents errors equal to zero.}
\label{fig:rhvsrini}
\end{figure*}

In Fig. \ref{fig:prof_fit}, we show an example of an extracted surface brightness profile along with the corresponding best-fit model. The corresponding residuals are also shown. In order to compare in a more homogeneous way the results obtained by {\sc nProFit}, we computed one of the most relevant quantities, the half-light radius ($\rm R_h$), and compared it with their initial values. In order to ensure that the obtained results are reliable, regarding the fitted model, we determined, the best fitting model in each case, following Eq. \ref{eq:delta_chi}, as we show in Fig. \ref{fig:delta_chi}. In panel (a), we show the comparison between the reference models and Moffat-EFF, and observe the overall trend of clusters initially modelled with Moffat-EFF not being well fitted in general by King and Wilson models. In panel (b), we show the comparison for clusters initially modelled with King models, resulting in good fits for Wilson fits, and poor fits for Moffat-EFF models. Finally, in panel (c), we show the comparison for clusters initially modelled with Wilson models, resulting in a similar behavior to that in the middle panel. Hence, from Fig. \ref{fig:delta_chi}, we conclude that clusters simulated with Moffat-EFF models are well fitted by Moffat-EFF models, whereas clusters simulated with King and Wilson models are well-fitted either by King or Wilson models. We computed the $\rm R_h$ values, from the fitted $r_0$ and $W_0$, for clusters best fitted either by King or Wilson models, and from $r_d$ and $\gamma$ for those best represented by Moffat-EFF models, following the prescription in Sec. \ref{sec:rh}.  We fit the three models to all the mock sample data, regardless of the models they were drawn from.

In order to test the accuracy of {\sc nProFit}, in Fig.~\ref{fig:rhvsrini}~(a), we show the difference between the initial $\rm R_h$ values and the values obtained by {\sc nProFit}, weighted by the initial $\rm R_h$ values, and compared with them, as a function of the difference between the central surface brightness and the measured background value ($\Delta \mu$). As expected, we observe a better recovery for larger $\Delta \mu$ values, even for small clusters, with larger errors, considering that the latter ones, have sizes close to the limit image resolution.

For the sake of this analysis, we removed extremely faint clusters in very crowded areas, resulting in fitting radius values shorter than 8 pixels, resulting in a sub-sample of 74 clusters. On average, we notice that the recovered values for clusters with $\rm \Delta\mu<5$~mag/arcsec$^2$ are within 20\% of the initial values, whereas, for clusters with $\rm \Delta\mu>5$~mag/arcsec$^2$, the recovered values are within 10\% of the initial values. 

In order to validate our results and compare them with the corresponding ones obtained by other publicly available tools, we carried out the structural parameters fitting using {\sc Galfit} \citep{Penggalfit} and {\sc Ishape} \citep{Larsenishape}. In  Fig.~\ref{fig:rhvsrini}~(b), we compare the results obtained by {\sc Galfit} (version 3.0.5), with the initial simulated values.  We obtained converging fits for 55 clusters, which are on average among the brightest ones.  We notice that {\sc Galfit} provides fits comparable to {\sc nProFit} in all size ranges.  On the other hand, in Fig.~\ref{fig:rhvsrini}~(c), we compare the results obtained by {\sc Ishape}, with the initial simulated values. For clusters with large contrast ($\Delta\mu\geq$5~mag/arcsec$^2$), {\sc Ishape} provides equally good results for 70\% of the clusters. However, the $\rm R_h$ values values obtained by {\sc Ishape} have larger dispersion and errors for $\Delta\mu<$5~mag/arcsec$^2$. \citet{Cuevas2020} had demonstrated that the core radii for real M82 clusters are well reproduced by {\sc Ishape}. Hence, the large error on $\rm R_h$ is most likely due to poor recovery of halo parameters in {\sc Ishape} for clusters located in high background regions.

We have also empirically found the $\rm R_h$ values by integrating the observed surface brightness profiles of each cluster (and corrected them by the PSF radius) and we compare them against the initial $\rm R_h$ values in Fig.~\ref{fig:rhvsrini}~(d). The recovery is especially poor for $\rm R_h<$4~pc, with the recovered values systematically larger.

\section{Conclusions}{\label{Sec:conclu}}

In this work, we present the numerical code {\sc nProFit}, devoted to obtain the best-fit structural parameters of star clusters in the HST images of nearby (distance $<5$~Mpc) galaxies.  The code is {\sc Python}-based at the user end, but uses modules of {\sc Pyraf} and {\sc Fortran}. {\sc nProFit} extracts sub-images centered in each analyzed object coordinates. Subsequently, a local background estimation is carried out by {\sc nProFit} determining the median values in the corners or by scanning the whole images and estimating the background value by a $\rm \sigma$-clipping procedure.  The estimated value is subsequently subtracted from the extracted SBPs by {\sc nProFit} from isophotal fittings.  PSF-convolved Moffat-EFF, King and Wilson models are then fitted to background subtracted azymuthally averaged surface brightness profiles.  {\sc nProFit} uses a $\rm \chi^2$-minimization technique to fit the models.  As a result, the tool provides the set of basic structural parameters, scale parameters ($\rm r_d$ for Moffat-EFF and $\rm r_0$ for King and Wilson models) and shape parameters ($\rm \gamma$ for Moffat-EFF and $\rm W_0$ for King and Wilson models).  Since nProFit fits dynamical models, it offers a valuable opportunity to derive physically-relevant parameters.  Among these parameters are central volume and luminosity densities ($\rm \rho_0$ and $\rm j_0$), total masses and luminosities ($\rm M$ and $\rm L$), central velocity dispersions ($\rm \sigma_0$), core radius ($\rm R_c$), half-light radius ($\rm R_h$), tidal radius ($\rm R_t$)  and binding energy ($\rm E_b$). 

We have tested {\sc nProFit} on simulated clusters superposed on real HST images. For the simulated clusters, the surface brightness difference between the cluster maximum and the local background varies between 3 to 8~mag/arcsec$^2$. We demonstrate that the input values are recovered within the 1-$\sigma$ errors for majority of the simulated clusters for all the three theoretical models we have explored. The $\rm R_h$ values are recovered
within 10~percent for clusters with $\Delta\mu>$5~mag/arcsec$^2$ and 20~percent for clusters with $\Delta\mu<$5~mag/arcsec$^2$. The accuracy of our recovery is comparable to that of {\sc Galfit}, whereas it is clearly better than that with {\sc Ishape}, especially for clusters $\Delta\mu<$5~mag/arcsec$^2$. We illustrate that {\sc nProFit} is a tool suitable to fit the structural parameters in samples with considerable crowding, such as the M82 disk, providing reliable values for clusters for which a neighbouring cluster does not contribute significantly within a distance of 8~HST/ACS pixels.

As a final note, we clarify that in this work, we considered that mass profiles follow light profiles over all clusters' radii. This is a simplification, since, mass segregation influences the mass density profiles of clusters, as well as their central velocity dispersions, causing the mass profiles to depart from the corresponding light profiles. For this reason, we will include mass segregation prescriptions in the upcoming version of  {\sc nProFit} .

%



\section*{Acknowledgments}
\addcontentsline{toc}{section}{Acknowledgements}

BCO thanks CONACyT for the support that enabled
her to carry out the work presented here. We also thank CONACyT for
the research grants CB-A1-S-25070 (YDM), CB-2014-240426 (IP), and
CB-A1-S-22784 (DRG), that allowed the acquisition of a cluster that
was used for computations in this work. BCO wants to thank the anonymous reviewer for the comments that allowed us to improve this work significantly.


\bibliography{main}{}
\bibliographystyle{aasjournal}

\appendix

\section{{\sc nProFit} Input}{\label{Ap:nprofit}}

We show an example of {\sc nProFit} input. {\sc nProFit} is a modular stand-alone code, fully adjustable to the user's needs. For further details on the routines see the README file.\\
\par\noindent
\begin{minipage}[c]{.2\textwidth}
\begin{lstlisting}[language=bash]
filters.dat
1 

100

list_x0_y0.dat
1
no
no
no
2
yes
no
mask_file.dat
yes
no
no
no
0.3
no
1
2

yes
yes
yes
nprofit_librarypath
3

yes
ds9_path
yes
\end{lstlisting}%
\end{minipage}%
\begin{minipage}[c]{.8\textwidth}
\begin{lstlisting}
#Filters information file
#Fitting box size option (1 same for all objects, 2 table with box size for 
each object)
#Box size (if the previous option is 1), Fitting box sizes information file 
(if the previous option is 2)
#Objects coordinates
#Coordinate system (1 Image(pixel), 2 WCS)
#Cut images into images centered in each object
#Substract sky
#Measure sky 
#Sky measurement option (1 for instat, 2 for median method, see README for details)
#Provide measurements of the sky in a file 
#Use pixel mask of contaminants?
#Pixel mask file (ASCII file)
#Use given ellipticity and P.A. for isophotal fitting (supplied by the user)
#Calculate P.A., ellipticity for isophotal fitting
#Calculate isophotal fitting
#Restrict ellipticity 
#Ellipticity restriction value (if previous option is 'yes')
#Convolve model with the PSF
#PSF options (1 user given, 2 Gaussian, etc.)
#Dynamical models fitting options (1 prepare data for fitting, 
2 data already prepared for fitting)
#Fit Moffat-EFF profile
#Fit King Dynamical profile
#Fit Wilson profile
#Absolute nProFit library path
#Fitting procedure options (1 automatically fit, 2 prepare a script to fit 
later the models, 3 fits already performed)
#Plot surface brightness profiles and dynamical models fitting
#DS9 path (if the previous option is yes)
#Compute derived parameters
\end{lstlisting}%
\end{minipage}
%

\end{document}